\documentclass[twoside,nohyper]{JHEP}
\usepackage{graphicx}

\newcommand{\ba}{\begin{eqnarray}}
\newcommand{\ea}{\end{eqnarray}}
\newcommand{\be}{\begin{equation}}
\newcommand{\ee}{\end{equation}}
\newcommand{\dis}{\displaystyle}
\newcommand{\tr}{\mbox{tr}}
\newcommand{\re}{\mbox{Re }}
\newcommand{\im}{\mbox{Im }}

\newcommand{\barr}{\begin{array}{c}}
\newcommand{\earr}{\end{array}}

\title{Matching the Electroweak Penguins \\ 
$Q_7$, $Q_8$ and Spectral Correlators} 
\author{ Johan Bijnens\\
Department of Theoretical Physics 2, Lund University\\
S\"olvegatan 14A, S 22362 Lund, Sweden}
\author{Elvira G\'amiz and Joaquim Prades\\
Centro Andaluz de F\'\i sica de las Part\'\i culas Elementales (CAFPE) and \\ 
Departamento de F\'\i sica Te\'orica y del Cosmos, 
Universidad de Granada,\\ Campus de Fuente Nueva, E-18002 Granada, Spain}

\abstract{Exact analytical expressions for 
the $\Delta S=1$ coupling $\im G_E$ in terms of observable spectral functions
are given.
This coupling determines the size of the $\Delta I=3/2$
contribution to $\varepsilon'$.
We show analytically how the scheme-dependence and scale dependences
vanish to all orders in $1/N_c$ and NLO in $\alpha_S$ explicitly
both for $Q_7$ and $Q_8$.

Numerical results are derived for both $Q_7$ and $Q_8$ from the
$\tau$-data and known results on the scalar spectral functions.
In particular we study the effect of all higher dimension operators.

The coefficients of the leading operators in the OPE of the needed
correlators are derived to NLO in $\alpha_S$.}

\preprint{LU TP 01-28\\CAFPE-2/01\\ UG-FT-131/01\\hep-ph/0108240\\revised
september 2001}

\begin{document}

\section{Introduction}

The lowest order SU(3) $\times$ SU(3)
 chiral Lagrangian describing $|\Delta S|=1$
transitions is given by
\ba
{\cal L}^{(2)}_{|\Delta S|=1}&=&
C\, F_0^6 \, e^2 \, G_E \, \tr \left( \Delta_{32} u^\dagger Q u\right)
+ C F_0^4 \left[ G_8 \, \tr \left( \Delta_{32} u_\mu u^\mu \right)
+ G_8' \tr \left( \Delta_{32} \chi_+ \right) \right.
\nonumber \\ 
&+& \left. 
G_{27} \, t^{ij,kl} \, \tr \left( \Delta_{ij} u_\mu \right) \,
\tr \left(\Delta_{kl} u^\mu\right) \right] + {\rm h.c.}
\ea
with
\be
C= -\frac{3}{5} \frac{G_F}{\sqrt 2} V_{ud} {V_{us}}^* \simeq -1.07 \cdot 
10^{-6} \, {\rm GeV}^{-2}
\ee
$F_0$ is the chiral limit value of the pion decay
constant $f_\pi= (92.4 \pm 0.4)$ MeV, 
\ba
u_\mu \equiv i u^\dagger (D_\mu U) u^\dagger = u_\mu^\dagger \; , 
\nonumber \\
\Delta_{ij}= u \lambda_{ij} u^\dagger\; (\lambda_{ij})_{ab}\equiv
\delta_{ia} \delta_{jb}
\ea 
and $U\equiv u^2=\exp{(i\sqrt 2 \Phi /F_0)}$ is the exponential
representation incorporating the octet of light pseudo-scalar mesons
in the SU(3) matrix $\Phi$.

The SU(3) $\times$ SU(3) tensor $t^{ij,kl}$ can be found
in \cite{BPP98} and $Q=\mbox{diag}(2/3,-1/3,-1/3)$ is a 3 $\times$ 3
matrix which collects the electric charge of the three light 
quark flavours.

In the Standard Model in the chiral limit, $\varepsilon'$ 
is essentially determined by the value of $(\im G_E)/ G_{27}$ and  
$(\im G_8)/(\re G_8)$ including Final State Interactions to  all
orders.  Therefore the knowledge of $\im G_E$ is of primordial importance.
The value of this parameter is dominated by the electroweak penguin
contributions $Q_7$ and $Q_8$ \cite{BW}.

The paper consists out of two parts. In the first part, Sections 
\ref{overview}--\ref{bag},  
we discuss how the $X$-boson approach takes care of the scheme-dependence
in the chiral limit independent of the large $N_c$ expansion we used
in our previous work. We also show precisely how the 
needed matrix-elements in the chiral limit
are related to integrals over spectral functions. This clarifies and extends
the previous work on this relation \cite{KPR99,DG00,NAR01,KPR01}.
Equation (\ref{imge}) is our main result, but we also present
the expression in terms of the usual bag parameters
in Section \ref{bag}.

In the second part, Sections \ref{PILR}--\ref{conclusions}, 
we present numerical results and compare our results
with those obtained by others and our previous work.
Sections \ref{PILR} and \ref{SSPP} describe
the experimental and theoretical information
on both $\im \Pi_{LR}^T(Q^2)$  and the 
scalar--pseudo-scalar 
$\im \Pi_{SS+PP}^{(0-3)}(Q^2)$ spectral functions and give the values
of the various quantities needed. The comparison with earlier results
is Section \ref{comparison}.

In addition, in the appendices we derive the NLO in $\alpha_S$
coefficient of the leading order term in the OPE of the needed
correlators in the same scheme as used for the short-distance weak
Hamiltonian. This coefficient was previously only known in a different
scheme \cite{LSC86}.

\section{Overview}
\label{overview}

This section describes the underlying reasoning elaborated in more
detail in the next two sections. In particular we use a simplified
notation here to allow simpler intermediate expressions, but we 
refer to the full
equations of the following sections.

We start from the effective action derived from the Standard Model
using short-distance renormalization group methods of
the form (Eq. (\ref{effective}))
\be
\label{step1}
\Gamma_{SD} = \sum_{i=7,8} C_i(\mu_R) Q_i(\mu_R)\,.
\ee
This effective action can be used directly in lattice calculations
but is less easy to use in other methods.
What we know how to identify are currents and densities. We therefore
go over to an equivalent scheme using only densities and currents
whereby we generate (\ref{step1}) by the exchange of colourless $X$-bosons
(Eq. (\ref{Xeffective}))
\be
\label{step2}
\Gamma_X = \sum_i g_i(\mu_C) X_i^I (\bar q' \gamma_I q)
\ee
where the coupling constants $g_i$ can be determined using short-distance
calculations only. The result is Eqs. (\ref{g7}) and (\ref{g8}).
At this step the scheme-dependence in the calculation of
the Wilson coefficients $C_i$
is removed but we have now a dependence on $M_X$ and the scheme used
to calculate with $\Gamma_X$.

We then need to evaluate the matrix-elements of (\ref{step2}).
For the case at hand this simplifies considerably. In the chiral limit,
the relevant matrix element can be related to  vacuum matrix elements (VEVs).
The disconnected contributions are just two-quark condensates.
The connected ones 
can be expressed as integrals over two-point functions (or correlators) as
given in Eq. (\ref{GE}), which we evaluate in Euclidean space.
The two relevant integrals are Eqs. (\ref{Q7}) and (\ref{Q8}).

Both of the integrals are now dealt with in a similar way.
We split them into two pieces at a scale $\mu$ via Eq. (\ref{split}).
The two-point function to be integrated over is replaced by its
spectral representation, which we assume known.

The long-distance part of the $Q^2$ integral can be evaluated
and integrals of the type (\ref{Q7LD})
and (\ref{Q8LD}) remain.

The short-distance part we evaluate in a somewhat more elaborate way
which allows us to show that the residual dependence on the 
$X$-boson mass disappears and that the correct behaviour given by
the renormalization  group is also incorporated.
To do this, we split the short-distance
integral in the part with the lowest dimensional operator,
which is of dimension six for both $Q_7$ and $Q_8$, 
and the remainder,  the latter is
referred to as the contribution from higher-order operators\cite{CDG00}.

The dimension six part can be evaluated using the known QCD 
short-distance 
behaviour of the two-point functions at this order. It is
vacuum expectation values of  dimension six operators over $Q^6$
for $Q_7$  and over $Q^4$ for $Q_8$
times a known function of $\alpha_S$. The vacuum expectation values
can be rewritten again as integrals over two-point functions
and the resulting integrals are precisely those needed to cancel
the remaining $M_X$-dependence.
For the contribution from {\em all} higher order operators we again
perform simply the relevant $Q^2$ integrals over the 
{\em same} two-point functions as for dimension six and they are 
the ones needed to match long- and short-distances exactly.

This way we see how our procedure precisely cancels  
all the scheme- and scale-dependence and fully relates the results to
known spectral functions.

\section{The $Q_7$ and $Q_8$ Operators}
\label{Q7Q8}
      
The imaginary part of $G_E$ is dominated by
the short-distance electroweak effects and can thus be reliably
estimated  from the purely strong matrix-elements of
the $|\Delta S|=1$ effective action below the charm quark mass 
\ba
\label{effective}
\Gamma_{{\rm eff}} &=& -\frac{G_F}{\sqrt 2} \, 
V_{ud} V_{us}^* \, {\dis \sum_{i=7,8}}  \, \im C_i \,\,
 \int {\rm d}^4 x \,  Q_i(x)
\ea
with $ \im  C_i =  y_i \, \im  \tau$ the imaginary part
of the Wilson coefficients, 
$\tau \equiv -\lambda_t/\lambda_u$ and
$\lambda_i \equiv V_{id}  V_{is}^*$. The $y_i$ coefficients are
known to two loops \cite{BJLW93,CFMR94} and 
\ba
Q_7&=& \left( \overline s_\alpha \gamma_\mu d_\alpha\right)_L
\, {\dis \sum_{q=u,d,s}} \frac{3}{2} e_q \left( \overline q_\beta
\gamma^\mu q_\beta \right)_R \,,
\\
Q_8&=& \left(\overline s_\alpha \gamma_\mu d_\beta \right)_L
\, {\dis \sum_{q=u,d,s}} \frac{3}{2} e_q \left( \overline q_\beta
\gamma^\mu q_\alpha \right)_R \, 
\ea
with $ \left( \overline q \gamma_\mu q \right)_{L(R)}
=\overline q \gamma_\mu (1-(+)\gamma_5) q$.
Up to $O(\alpha_S^2)$, the $Q_7$ and $Q_8$ operators
only mix  between themselves below the charm quark mass
via the strong interaction.

 The QCD anomalous dimension matrix $\gamma(\nu)$ 
in regularizations like Naive Dimensional Regularization (NDR)
or 't Hooft-Veltman (HV) which do not mix
operators of different dimension, is defined as\footnote{In a cut-off
regularization one has, on the right hand side, an infinite
series of higher dimensional operators suppressed by powers of the cut-off.
Explicit expressions for the matrices $\gamma^{ji}(\nu)$
are in App. \ref{AppA}.}
($i=$ 7,8)
\ba
\label{gamma}
\nu \, \frac{{\rm d}}{{\rm d} \nu } Q_i(\nu) &=&
- {\dis \sum_{j=7,8}} \gamma^{ji}(\nu) \, Q_j(\nu) \, ; \quad\quad
\gamma(\nu)=\sum_{n=1} \gamma^{(n)}a^n(\nu) \, 
\ea
where $a(\nu) \equiv \alpha_S(\nu) /\pi$.

At low energies, 
it is convenient to describe the $\Delta S=1$ transitions
with an effective action $\Gamma_{LD}$ which
uses hadrons, constituent quarks, or other objects
to describe the relevant degrees of freedom.
A four-dimensional regularization scheme like an Euclidean cut-off,
separating long-distance physics from integrated out short-distance
physics, is also more practical. In addition, the 
color singlet Fierzed 
operator basis becomes  useful for identifying QCD currents 
and densities.
The whole procedure has been explicitly done in \cite{scheme,epsprime}
and reviewed in \cite{talks,benasque}.

At low energies, the effective action (\ref{effective}) is therefore 
replaced by the equivalent
\ba
\label{Xeffective}
\Gamma_X&=& \ g_7(\mu_C,\cdots)
  X_7^\mu \, \left( (\overline s \gamma_\mu d )_L
 + \frac{3}{2} e_q 
{\dis \sum_{q=u,d,s}} (\overline q \gamma_\mu q )_R \right) 
\nonumber  \\  &+& 
g_8(\mu_C,\cdots) {\dis \sum_{q=u,d,s}} X_{q,8} \left( 
(\overline q d)_L 
+ (-2) \frac{3}{2} e_q (\overline s q)_R 
\right) \, .
\ea
Here all colour sums are performed implicitly inside the brackets.
There is also a kinetic term for the X-bosons which we take
to be all of the same mass for simplicity. 

The couplings $g_i$ are determined as functions of the 
Wilson coefficients $C_i$ by taking matrix elements of both sides
between quark and gluon external states as explained in 
\cite{scheme,epsprime,talks,benasque}. We obtain
\ba
\label{g7}
\frac{|g_7(\mu_C)|^2}{M_X^2}&=&
\im C_7(\mu_R) \, \left[ 1 + a(\mu_C) \, \left( 
\gamma^{(1)}_{77} \ln \frac{M_X}{\mu_R} + \Delta r_{77}  \right)
 \right]
 \nonumber \\
&+&   \im C_8(\mu_R) \,  \left[ a(\mu_C) \,  \Delta r_{78}  \right]
+ O \big( a(\mu_R)-a(\mu_C) \big)  
\ea
and
\ba
\label{g8}
\frac{|g_8(\mu_C)|^2}{M_X^2}&=&
\im C_8(\mu_R) \, \left[ 1 + a(\mu_C) \, \left( 
\gamma^{(1)}_{88} \ln \frac{M_X}{\mu_R}  
+ \tilde \gamma^{(1)}_{88} \ln \frac{\mu_C}{M_X} + 
\Delta r_{88}  \right) \right] \nonumber \\
&+& \im C_7(\mu_R) \,  \left[ a(\mu_C) \, \left( 
\gamma^{(1)}_{87} \ln \frac{M_X}{\mu_R} + \Delta r_{87}  \right)
 \right]
+ O \big( a(\mu_R)-a(\mu_C) \big) \, . \nonumber \\
\ea
 $\tilde \gamma_{ij}^{(1)}$ is due to the anomalous
 dimensions of the two-quark color-singlet  
densities or currents.  It vanishes for conserved currents.
In our case $\tilde \gamma_{88}^{(1)}= -2 \gamma_m^{(1)}$,
where $\gamma_m^{(1)}$ is the QCD anomalous dimension of the quark mass
in the regularization used in (\ref{Xeffective}).
The values of $\Delta r_{ij} \equiv (r-\tilde r)_{ij}$  
have been calculated in \cite{epsprime}. 

The effective action to be used at low-energies is now specified 
completely.
Notice that singlet color currents and densities are connected
by the exchange of a colourless X-boson and therefore are well
identified also in the low energy effective theories, and
the finite terms which appear
due to the correct identification of currents and densities.

The coupling $G_E$ is defined in the chiral limit so that
we can use soft pion theorems to calculate the relevant matrix-elements,
and relate them to a vacuum-matrix-element\footnote{In the real 
$K\to\pi\pi$
case we would need to evaluate integrals over strong-interaction five
point functions, three meson legs and two $X$-boson legs. For
vacuum matrix-elements this reduces to integrals over 
two-point functions, the two $X$-boson legs.
The same is not possible for $G_8$  and $G_{27}$
since the corresponding terms are order $p^2$
and have zero vacuum matrix elements.}.
For the contribution of $Q_7$ and $Q_8$, we obtain 
\ba
\label{GE}
-\frac{3}{5} \, 
e^2 \, F_0^6 \, \im G_E &=&  -|g_7(\mu_C, \cdots)|^2 \, 3 \, 
\, i \, \int \frac{{\rm d}^4 p_X}{(2\pi)^4} \, \frac{1}{p_X^2-M_X^2}\, 
g_{\mu \nu} \, \Pi_{LR}^{\mu\nu}(p_X^2) \nonumber \\
&+&    |g_8(\mu_C, \cdots)|^2  \, 
i \, \int \frac{{\rm d}^4 p_X}{(2\pi)^4} \, \frac{1}{p_X^2-M_X^2}\, 
\left( \Pi_{SS+PP}^{(0)}(p_X^2)- \Pi_{SS+PP}^{(3)}(p_X^2) \right)
 \, . \nonumber \\
\ea
Where $\Pi_{LR}^{\mu\nu}(p^2)$ is 
the following two-point function in the chiral limit 
\cite{KPR99,DG00}:
\ba
\Pi_{LR}^{\mu\nu}(p)&\equiv&
\frac{1}{2} \, i \, \int {\rm d}^4 y \, e^{i y \cdot p } \, 
\langle 0| T \left[ L^\mu(y) R^{\nu\dagger}(0) \right] | 0 \rangle
 \equiv \left[ p^\mu p^\nu -g^{\mu\nu} p^2\right] \Pi_{LR}^T(p^2)
\nonumber \\ 
&+& p^\mu p^\nu  \Pi_{LR}^L(p^2)\,.
\ea 
In Eq. (\ref{GE}) we used the chiral limit
so  SU(3) chiral symmetry is exact.
$L(R)^\mu= (\overline u \gamma^\mu  d)_{L(R)}$
or  $L(R)^\mu= (\overline d \gamma^\mu  s)_{L(R)}$,
$\Pi^L_{LR}(p^2)$ vanishes
and $\Pi_{SS+PP}^{(a)}(p^2)$ 
is the two-point function
\ba
\Pi_{SS+PP}^{(a)}(p^2)&\equiv&
i \, \int {\rm d}^4 y \, e^{i y \cdot p} \, 
\langle 0 | T \left[ (S+iP)^{(a)}(y) (S-iP)^{(a)}(0) \right] | 0 \rangle
\ea
with  
\ba
S^{(a)}(x)&=& -\overline q(x) \frac{\lambda^{(a)}}{\sqrt 2} q (x), \, \, 
P^{(b)}(x)= \overline q(x) i \gamma_5 
\frac{\lambda^{(a)}}{\sqrt 2} q (x)\,.
\ea
The 3 $\times$ 3 matrix 
$\dis\lambda^{(0)} = \sqrt{2}\,{I}/{\sqrt 3}$ 
and the rest are the Gell-Mann matrices normalized to
$
\tr \left( \lambda^{(a)}  \lambda^{(b)} \right) = 2 \delta^{ab}  
$.
An alternative form for the last term in (\ref{GE}) is,
\be
\label{alternative}
   |g_8(\mu_C, \cdots)|^2  \, 
3 \, i \, \int \frac{{\rm d}^4 p_X}{(2\pi)^4} \, \frac{1}{p_X^2-M_X^2}\, 
 \Pi_{SS+PP}^{(ds)}(p_X^2)  
\ee
with 
\be
 \Pi_{SS+PP}^{(ds)}(p^2) = 
i \, \int {\rm d}^4 y \, e^{i y \cdot p} \, 
\langle 0 | T \left[ (\overline d d)_L (y)
          (\overline s s)_R(0) \right] | 0 \rangle
\ee
and $ \left( \overline q  q \right)_{L(R)}
=\overline q (1-(+)\gamma_5) q$.

\section{Exact Long--Short-Distance Matching at NLO in $\alpha_S$}
\label{exactmatching}

\subsection{The $\mathbf{Q_7}$ contribution}
In Euclidean space, the term multiplying $|g_7|^2$ 
in the rhs of (\ref{GE}) can be written as
\ba
\label{Q7}
-\frac{9}{16\pi^2} \int^\infty_0 {\rm d} Q^2 \frac{Q^4}{Q^2+M_X^2}
\, \Pi_{LR}^T(Q^2)
\ea
with $Q^2=-q^2$.
We split the integration into a short-distance
and a long-distance part by
\ba
\label{split}
\int^{\infty}_0 \, {\rm d} Q^2  &=&
 \int^{\mu^2}_0 \, {\rm d} Q^2  +
 \int^{\infty}_{\mu^2} \, {\rm d} Q^2 
\ea
with $M_X^2 >>\mu^2$. 
In QCD, $\Pi_{LR}^T(Q^2)$ obeys an unsubtracted dispersion relation
\ba
\label{KL}
\Pi_{LR}^T(Q^2)&=& \int_0^\infty    
 {\rm d} t \, \frac{1}{\pi} 
\frac{\im \Pi_{LR}^T(t)}{t+Q^2} \, .
\ea  

\subsubsection{$\mathbf{Q_7}$ Long-distance}

Putting (\ref{KL}) in (\ref{Q7}) and performing the integral
up to $\mu^2$ gives
\be
\label{Q7LD}
-\frac{9}{16\pi^2} \int_0^\infty dt
\frac{t^2}{M_X^2} \ln\left(1+\frac{\mu^2}{t}\right)
 \frac{1}{\pi}{\im \Pi_{LR}^T(t)}+{\cal O}
\left(\frac{\mu^2}{M_X^4}\right)\,,
\ee
with the use of the Weinberg Sum Rules
\cite{WEIN67}, Eqs. (\ref{WSRS}).

\subsubsection{$\mathbf{Q_7}$ Short-distance}

At large $Q^2$ in the chiral limit, $\Pi_{LR}^T(Q^2)$ 
behaves in QCD as \cite{SVZ79} 
\ba
\label{SVZLR}
\Pi_{LR}^T(Q^2)&\to& 
\sum_{n=0}^\infty \,
\sum_{i=1} \,
 \frac{ C^{(i)}_{2(n+3)}(\nu, Q^2)}{Q^{2(n+3)}} \, 
\langle 0|  O^{(i)}_{2(n+3)}(0) | 0 \rangle (\nu)
\ea
where $O^{(i)}_{2(n+3)}(0)$ are dimension $2(n+3)$ 
gauge invariant operators.
\ba
\label{D6_1}
 O^{(1)}_6(0) &=&    \frac{1}{4} \, L^\mu(0) \, R_\mu(0) =
 \frac{1}{4} \left( \overline s \gamma^\mu  d \right)_L (0)
\left( \overline d \gamma_\mu  s \right)_R (0) \, ; 
\nonumber\\
 O^{(2)}_6(0) &=&  
(S+iP)^{(0)}(0) \, (S-iP)^{(0)}(0) 
- (S+iP)^{(3)}(0) \, (S-iP)^{(3)}(0)  \nonumber \\
&=& 3 \, (\overline d d)_L (0)
(\overline s s)_R (0)\,.
\ea
The coefficients $C_6^{(i)}(\nu,Q^2)$ are related to
the anomalous dimension matrix defined in (\ref{gamma}).
This can be used to obtain the NLO in $\alpha_S$ part of the coefficient
with the same choice
of evanescent operators as in \cite{BJLW93,BBL96},
calculations of the $\alpha_S^2$ term in other schemes and choices of 
evanescent operators are in\cite{LSC86}.
Our calculation and results are in App. \ref{AppA}.
At the order we work
we only need the
lowest order \cite{SVZ79}
\ba
\label{D6_2}
C^{(1)}_{6}(\nu,Q^2)&=& 
-\frac{16 \pi^2 a(\nu)}{3}
%\left\{ 
\gamma_{77}^{(1)}
%+ a(\nu)
%\left[ \gamma_{77}^{(1)}+
%\gamma_{77}^{(2)} + \frac{\gamma_{77}^{(1)}}{2} 
%\left( \beta_1 - \gamma_{77}^{(1)}\right) 
%\left(\ln \left( \frac{Q^2}{\nu^2} \right) -\frac{1}{3} 
%\right) \right]  \right\} \, ; 
\nonumber \\
C^{(2)}_{6}(\nu,Q^2)&=&  
\frac{8 \pi^2 a(\nu)}{9}
%\left\{
\gamma_{87}^{(1)} 
%+ a(\nu)  
%\left[ \gamma_{87}^{(1)}+\gamma_{87}^{(2)}  +  \frac{\gamma_{87}^{(1)}}{2}
% \left( \beta_1 - \gamma_{88}^{(1)} - \gamma_{77}^{(1)} \right)
%\left(\ln \left( \frac{Q^2}{\nu^2} \right) -\frac{1}{3}
%\right)\right] \right\}  .
\ea

The values of the coefficients of the power corrections 
are physical quantities and can be determined
with global duality
FESR\footnote{The specific form (\ref{FESR}) is only true to lowest
order in $\alpha_S$ due to the $\ln(Q^2)$ dependence at higher orders.},
 \cite{ALEPH98,OPAL99,FESRtalks},
\ba
\label{FESR}
{\dis \sum_{m=0}^\infty \, 
\sum_{i=1}} (-1)^m \, 
\langle 0| O^{(i)}_{2(m+3)}(0) | 0 \rangle(s_0) \, 
\frac{1}{2 \pi i} \, \oint_{C_{s_0}} \,{\rm d} s \, 
\frac{C^{(i)}_{2(m+3)}(s_0,-s)}{s^{1+m-n}} 
\nonumber \\ = M_{n+2}\equiv  \int_0^{s_0} {\rm d} t \, t^{n+2} \, 
\frac{1}{\pi}  \, \im \Pi_{LR}^T(t) \, , 
\ea
with $n \geq 0$. $s_0$ is the threshold for local
duality\footnote{A discussion of the value of the local duality onset
is in Section  \ref{PILR}.}.
At leading order in $\alpha_S$ only $n=m$ survive
and we can rewrite the short-distance 
contribution to (\ref{Q7}) as 
\ba
\label{shortQ7}
\lefteqn{
-\frac{9}{16\pi^2} \int^\infty_{\mu^2} {\rm d} Q^2 \frac{Q^4}{Q^2+M_X^2}
\, \Pi_{LR}^T(Q^2) =}&&\nonumber\\
&=&a(\mu) \, \ln \frac{\mu}{M_X}  
 i \int \frac{{\rm d}^4 \tilde q}{(2\pi)^4}  \, 
\left( \frac{1}{M_X^2} \right) \, \,  
\left[ \gamma^{(1)}_{77} \, 3 g_{\mu\nu} \, \Pi_{LR}^{\mu\nu}(\tilde q) -
 \gamma^{(1)}_{87} \,
\left( \Pi_{SS+PP}^{(0)}(\tilde q^2)- 
\Pi_{SS+PP}^{(3)}(\tilde q^2) \right) 
\right] \nonumber \\ 
&&+  \left(\frac{-1}{M_X^2}\right)
\frac{9}{16 \pi^2} \, \sum_{n=1}^{\infty} \frac{1}{n} \,  
\sum_{i=1} \frac{ C^{(i)}_{2(n+3)}}{\mu^{2n}} 
\left[ 1 + O\left( \frac{\mu^2}{M_X^2} \right) \right]
\langle 0| O^{(i)}_{2(n+3)}(0) | 0 \rangle + O(a^2) + \cdots
\nonumber \\
&=&  a(\mu)\, \ln \frac{\mu}{M_X}  
 i \int \frac{{\rm d}^4 \tilde q}{(2\pi)^4}  \, 
\left( \frac{1}{M_X^2} \right) \, 
\, \left[ \gamma^{(1)}_{77} 
\, 3g_{\mu\nu} 
\, \Pi_{LR}^{\mu\nu}(\tilde q) - \gamma^{(1)}_{87} \,
\left( \Pi_{SS+PP}^{(0)}(\tilde q^2)- 
\Pi_{SS+PP}^{(3)}(\tilde q^2) \right) 
\right]  \nonumber \\ 
&&+  \frac{9}{16 \pi^2} \, 
\int_0^{s_0} {\rm d} t \, 
\frac{t^2}{M_X^2} \, \ln \left( 1+ \frac{t}{\mu^2} \right) 
\, \frac{1}{\pi}  \, \im \Pi^T_{LR}(t) 
 + {\cal O}\left( \frac{\mu^2}{M_X^4} \right)
+ {\cal O}(a^2)
\ea
where we have used
\ba
\label{Tordered}
 \int \frac{{\rm d}^4 \tilde q }{(2\pi)^4}  \, 
\int {\rm d}^4 x  \, e^{i x\cdot\tilde q}
\, \langle 0|  T \left[ J(x) \, \tilde J(0) \right] | 0 \rangle
&\equiv& \langle 0|  J(0) \, \tilde J(0) | 0 \rangle \, .  
\ea

\subsection{The $\mathbf{Q_8}$ Contribution}

In Euclidean space,  
the term multiplying $|g_8|^2$  in the rhs of (\ref{GE}), is
\ba
\label{Q8}
\frac{1}{16 \pi^2} \int^\infty_{0} 
{\rm d} Q^2 \frac{Q^2}{Q^2+M_X^2} \, \Pi_{SS+PP}^{(0-3)}(Q^2) \, 
 \ea
with
\ba
 \Pi_{SS+PP}^{(0)}(Q^2)- \Pi_{SS+PP}^{(3)}(Q^2) 
&\equiv&  \Pi_{SS+PP}^{(0-3)}(Q^2) \, . 
\ea
This two-point function has a disconnected contribution,
corresponding to what is usually called the factorizable
contribution\footnote{For the other operators this
correspondence does not hold and even for $Q_8$
it is only valid in certain schemes, including ours.}.
We split off that part explicitly:
\ba
\label{Q8conn}
 \frac{1}{16 \pi^2}
\int^\infty_{0} {\rm d} Q^2 \frac{Q^2}{Q^2+M_X^2}
\, \Pi_{SS+PP}^{(0-3)}(Q^2) \, 
= \frac{1}{M_X^2} \,  \big| \langle 0 | S^{(0)}(0) | 0 \rangle \big|^2 \,  
\nonumber \\ +  \frac{1}{16 \pi^2}
\int^\infty_{0} {\rm d} Q^2 \frac{Q^2}{Q^2+M_X^2}
\, \Pi_{SS+PP}^{(0-3)~conn}(Q^2) \,. 
\ea 

\subsubsection{The Disconnected Contribution}
\label{disconnected}

We have included in (\ref{g8})
all the $O(\alpha_S)$ logs and finite terms 
that take into account passing the four-quark
matrix element from the cut-off $\mu_C$ 
regulated $X$-boson effective theory to the $\overline{MS}$ one.
Therefore to the order needed
\be 
\langle 0 | S^{(0)}(0) | 0 \rangle^2 = 3 \, 
\langle 0 | \overline q q | 0 \rangle \big|^2_{\overline{MS}}(\mu_C)
\ee
and from now on the quark condensate is understood to be in
the $\overline{MS}$ scheme.
As shown in \cite{BG86,deR89}, $\gamma_{88}^{(1)} = - 2 \gamma_m^{(1)}$
where $\gamma_m^{(1)}$ is the one-loop quark mass anomalous
dimension\footnote{ See App. \ref{AppA} for the explicit expressions.
It can be seen there that no such relation holds
for $\gamma_{88}^{(2)}$.}. 
This cancels exactly the scale $\mu_C$ dependence in (\ref{g8})
to order $\alpha_S$\cite{BG86,deR89}.

The disconnected contribution to $\im G_E$ is thus
\be
\label{Q8disc}
-\frac{3}{5} \, e^2 F_0^6 \, \im G_E^{Fact} = 
 3 \, \langle 0 | \bar q q | 0 \rangle ^2 (\mu_C)
 \frac{|g_8(\mu_C)|^2}{M_X^2}
\ee
but now with
\ba
\frac{|g_8(\mu_C)|^2}{M_X^2}
&=&  \im C_8(\mu_R) 
\left[ 1 + a(\mu_C)  \left( \gamma^{(1)}_{88}
\ln \frac{\mu_C}{\mu_R}  + \Delta r_{88}  
\right) \right] 
\nonumber \\
&+&   \im C_7(\mu_R) a(\mu_R)  
\left( \gamma^{(1)}_{87}
\ln \frac{M_X}{\mu_R}  + \Delta r_{87} 
\right)
+  {\cal O}(a^2)\,.
\ea
Here one can see that the factorizable contribution
is not well defined. It is due to the mixing of $Q_7$
and $Q_8$ and  is reflected here in the $\ln(M_X/\mu_R)$.
This $M_X$ dependence  cancels with
the {\em non-factorizable} contribution of $Q_7$ 
in (\ref{shortQ7}). Notice
that the contribution of both terms, $\im C_8$  and $\im C_7$, 
to $ \im G_E$ are of the same order in $1/N_c$. 
 It is then necessary to add the non-factorizable
term  to have $\im G_E$  well defined.
Since $\im G_E$ is a physical quantity,
factorization is \underline{not} well defined for $Q_8$.
This was also shown to be the case for $Q_6$ in  \cite{deltaI}. 
Of course, the leading term of the $1/N_c$
expansion is well defined but that approximation
would  miss a completely new topology, namely
the non-factorizable contributions.

\subsubsection{The Connected Contribution}
\label{connected}

{}From the leading high energy behaviour,
the scalar--pseudo-scalar spectral functions satisfy
in the chiral limit \cite{GL84,BDLW75}
\be
\label{weinSP1}
\int_0^{\infty} dt\, \frac{1}{\pi}  \, 
\left[ \im \Pi_{SS}^{(0)}(t) - \im \Pi_{PP}^{(3)}(t) \right]
= 0 =
\int_0^{\infty} dt\, \frac{1}{\pi}  \, 
\left[ \im \Pi_{SS}^{(3)}(t) - \im \Pi_{PP}^{(0)}(t) \right]  
\ee
which are analogous to  Weinberg Sum Rules.
Therefore the connected part of  
$\Pi^{(0-3)}_{SS+PP}(Q^2)$
satisfies an unsubtracted dispersion relation in the chiral limit,
\be
\label{dispersionSS}
\Pi^{(0-3) \,\rm conn}_{SS+PP}(Q^2)= 
{\dis \int^\infty_0} {\rm d}t  \, \frac{1}{\pi} \, 
\frac{\im \Pi^{(0-3)}_{SS+PP}(t)}{t+Q^2} \, .
\ee

Also in the chiral limit, the scalar and pseudo-scalar $(0-3)$ 
combinations satisfy other Weinberg-like Sum Rules 
as shown in \cite{MOU00} for the scalar\footnote{In \cite{MOU00} it
was the alternative form of Eq. (\ref{alternative}) which was 
used.}   and in \cite{LEU90} for the pseudo-scalar, 
\be
\label{weinSP2}
\int_0^{\infty} dt\, \frac{1}{\pi}  \,  \im \Pi_{SS}^{(0-3)}(t)= 0 =
\int_0^{\infty} dt\, \frac{1}{\pi}  \,  \im \Pi_{PP}^{(0-3)}(t)\,.
\ee

We also know that the spectral functions $\im \Pi_{SS(PP)}(Q^2)$
depend on scale due to the non-conservation of the quark densities. 
\ba
\label{scaleSS}
\mu_C \frac{{\rm d} }{{\rm d} \mu_C } \,  
\im \Pi_{SS(PP)}^{(a)}(t)
 &=&  2 \gamma_m(\mu_C) \, \im \Pi_{SS(PP)}^{(a)}(t)  
\, .
\ea
This scale dependence is analogous to the one of the disconnected
part (\ref{Q8disc}) 
and  cancels the $\mu_C$ dependence in $|g_8(\mu_C)|^2$ 
also for the connected part.

We now proceed as for  $Q_7$ and split the
integral in (\ref{Q8conn}) at $\mu^2$.

\subsubsection*{\ref{connected}.a $\mathbf{Q_8}^{conn}$ Long-Distance}

We perform simply the integral and obtain
\be
\label{Q8LD}
-\frac{1}{16\pi^2} \int_0^\infty dt
\frac{t}{M_X^2} \ln\left(1+\frac{\mu^2}{t}\right)
 \frac{1}{\pi}{\im \Pi_{SS+PP}^{(0-3)}(t)}+{\cal O}
\left(\frac{\mu^2}{M_X^4}\right)\,.
\ee

\subsubsection*{\ref{connected}.b $\mathbf{Q_8}^{conn}$
Short-distance}

Using the unsubtracted dispersion relation in (\ref{dispersionSS}),
$\Pi_{SS+PP}^{(0-3) \, \rm conn}(Q^2)$ in the chiral limit
behaves at  large $Q^2$  in QCD as
\ba
\label{SVZSP}
\Pi_{SS+PP}^{(0-3)\, \rm conn}(Q^2) &\to& 
\sum_{n=0}^\infty \, \sum_{i=1} \,
 \frac{ \tilde C^{(i)}_{2(n+3)}(\nu, Q^2)}{Q^{2(n+2)}} \, 
\langle 0|  \tilde O^{(i) \rm}_{2(n+3)}(0) | 0 \rangle
 (\nu) 
\ea
where $\tilde O^{(i)}_{2(n+3)}(0)$ are dimension $2(n+3)$ 
gauge invariant operators.
\be
 \tilde O^{(1)}_6(0) =    O^{(1)}_6(0)\, ; \quad\quad
 \tilde O^{(2)}_6(0) =  O^{(2)}_6(0)  \,.
\ee

Using the information on the  mixing of $Q_7$ and $Q_8$ 
in (\ref{gamma}), the scale dependence  (\ref{scaleSS}),
it is easy to obtain the leading power behavior in (\ref{SVZSP})
(see Appendix \ref{AppB})
\ba
\label{M1sumrule}
\tilde C^{(1)}_{6}(\nu,Q^2)&=& 
\frac{45\pi^2}{2}  a(\nu)^2  + O(a^3)  
\, ; \nonumber \\
\tilde C^{(2)}_{6}(\nu,Q^2)&=&  
 \frac{211\pi^2}{4}  a(\nu)^2  + O(a^3)  \, .
\ea

Again the values of the coefficients of the power corrections
in (\ref{SVZSP}) can be calculated using global duality FESR, 
\ba
{\dis \sum_{m=0}^\infty \, \sum_{i=1}} (-1)^{m+1}
\langle 0| \tilde O^{(i)}_{2(m+3)}(0) | 0 \rangle(\tilde s_0) \, 
\frac{1}{2 \pi i} \, \oint_{C_{\tilde s_0}} {\rm d} s \,  
\frac{\tilde C^{(i)}_{2(m+3)}(\tilde s_0,-s)}{s^{1+m-n}}
 \nonumber \\ =
\tilde M_{n+1} \equiv 
\int_0^{\tilde s_0} {\rm d} t \, t^{n+1} \, \frac{1}{\pi}  \, 
 \im \Pi^{(0-3)}_{SS+PP}(t)  \, ,
\ea
with $n\geq 0$, and $\tilde s_0$ the  threshold for local duality
for this two-point function.
Again only terms with $n=m$ survive at $O(\alpha_S)$
and one gets
\ba
\label{shortQ8}
\lefteqn{\frac{1}{16 \pi^2}
\int^\infty_{\mu^2} {\rm d} Q^2 \frac{Q^2}{Q^2+M_X^2}
\, \Pi_{SS+PP}^{(0-3)~conn}(Q^2)=}&&
\nonumber \\ 
&&\frac{1}{16 \pi^2} \,  
\int_0^{\tilde s_0} {\rm d} t \, 
\frac{t}{M_X^2} \, 
\ln \left( 1+ \frac{t}{\mu^2} \right)  \, 
\frac{1}{\pi}  \, \im \Pi^{(0-3)}_{SS+PP}(t)
 + {\cal O}\left( \frac{\mu^2}{M_X^4} \right) 
+ {\cal O}(a^3).
 \ea

\subsection{Sum}

We now  add all the contributions of Eqs. (\ref{Q7LD}),
(\ref{shortQ7}), (\ref{Q8disc}), (\ref{Q8LD})
and (\ref{shortQ8}) to obtain the full result.
Notice in particular that all contributions contain
the correct logarithms of $M_X$ to cancel that dependence
in Eqs. (\ref{g7}) and (\ref{g8}).

The integrals over the spectral functions in the respective long
and short-distance regime can in both cases be combined to
give a simple $\ln(t/\mu^2)$.

Therefore, when summing everything to ${\cal O}(\alpha_S)$ 
and {\em all} orders in $1/N_c$, we obtain
\ba
\label{imge}
-\frac{3}{5} \, e^2 F_0^6 \im G_E &=& 
   \Bigg\{ \im C_7(\mu_R) 
\left[1+ a (\mu_C)
\left( \gamma^{(1)}_{77} \ln \frac{\mu}{\mu_R} + \Delta r_{77}  
\right)\right] 
\nonumber \\
 &&+   \im C_8 (\mu_R) a(\mu_C)
 \Delta r_{78}  \Bigg\}   \frac{9}{16\pi^2}   {\cal A}_{LR}(\mu) \nonumber \\ 
&+&  \left\{\im  C_8(\mu_R) \left[1+ a(\mu_C)
\left( \gamma^{(1)}_{88} \ln \frac{\mu_C}{\mu_R}
 + \Delta r_{88}  \right) \right] \right. 
\nonumber \\
&&+ \left. \im C_7(\mu_R) a(\mu_C) 
\left( \gamma^{(1)}_{87} \ln \frac{\mu}{\mu_R} + 
\Delta r_{87}\right) \right\} \times
\nonumber \\
&& \quad\times
\left( 3 \,  \langle 0 | \bar q q | 0 \rangle^2(\mu_C) 
+ \frac{1}{16\pi^2} {\cal A}_{SP}(\mu,\mu_C)) \right)\, ;
\ea
where
\ba
\label{ALRSP}
{\cal A}_{LR}(\mu) &\equiv&
 \int_0^{s_0} {\rm d} t \, t^2 \, \ln \left(\frac{t}{\mu^2} \right) 
\, \frac{1}{\pi}  \, \im \Pi^T_{LR}(t) \, ; \nonumber \\
{\cal A}_{SP}(\mu,\mu_C) &\equiv&
\int_0^{\tilde s_0}{\rm d}t \, t
 \ln\left(\frac{t}{\mu^2}\right)
\frac{1}{\pi}\im \Pi_{SS+PP}^{(0-3)}(t) \, .
\ea

To obtain this result we have used the local duality relations
\ba
\int^{\infty}_{s_0} {\rm d} t \, t^2 \, 
\ln \left(\frac{t}{\mu^2} \right) 
\, \frac{1}{\pi}  \, \im \Pi^T_{LR}(t) &=& O(a^2)  \, , \nonumber  \\
\int^{\infty}_{\tilde s_0} {\rm d} t \, t \, 
\ln \left(\frac{t}{\mu^2} \right) 
\, \frac{1}{\pi}  \, \im \Pi^{(0-3)}_{SS+PP}(t) &=&O(a^2)  \, .
\ea

The expression in (\ref{imge})
 is exact in the chiral limit, at NLO in $\alpha_S$,
all  orders in $1/N_c$
and without electromagnetic corrections. 
It doesn't depend on any scale nor scheme
at that order analytically.  
The dependence on $M_X$ also nicely cancels out.
The $\mu_C$ dependence cancels against the $\mu_C$
dependence of the densities.
Notice that in this final result
we have taken into account the contribution
of all  higher order operators.

This result differs from the ones  in  
\cite{KPR99,DG00,NAR01,KPR01} in the finite terms 
$\Delta r_{88}$ and $\Delta r_{78}$. They are necessary
to cancel the scheme dependence. In addition,
\cite{DG00,NAR01} only take into account the dimension eight
corrections and \cite{KPR01} uses a hadronic large $N_c$ Ansatz
to estimate them.

As noticed in \cite{epsprime},  the connected
scalar--pseudo-scalar two-point function is
exactly zero in U(3)  symmetry, i.e.  is  $1/N_c$ suppressed. 
We used this fact to disregard this contribution there.
We will check later the quality of this approximation from a 
phenomenological analysis of its value.

\section{Bag Parameters}
\label{bag}

We now re-express our main result (\ref{imge}) in terms of the
usual definition of the bag parameters
\ba
-\frac{3}{5} \, e^2 F_0^6 \im G_E &\equiv& 
\langle 0 | \bar q q | 0 \rangle^2(\mu_C)
\left[ \im C_7(\mu_R) B_{7\chi}(\mu_C,\mu_R) + 
3 \im C_8(\mu_R) \, B_{8\chi}(\mu_C,\mu_R) \right]
\nonumber \\
&\equiv&  - 6 \, \im C_7(\mu_R) \,  
\langle 0 | O_6^{(1)} | 0 \rangle_\chi (\mu_R) +
\im C_8(\mu_R) \,  \langle 0 | O_6^{(2)} | 0 \rangle_\chi (\mu_R) \, ,
\ea
where the subscript $\chi$ means in the chiral limit.

This definition coincides with the one in \cite{epsprime}
and gives 
\ba
B_{7\chi}
(\mu_C,\mu_R) &=& \left[ 1 +  \Delta r_{77} \, a(\mu_R) \right]  
\, \frac{9}{16 \pi^2} \, 
\frac{1}{\langle 0 | \bar q q | 0 \rangle^2(\mu_C)}\, 
{\cal A}_{LR}(\mu_R) \nonumber \\ 
&+& 3 \,  a(\mu_C) \, \Delta r_{87} 
\left[ 1 + \frac{1}{48 \pi^2} \, 
\frac{1}{\langle 0 | \bar q q | 0 \rangle ^2(\mu_C)}
\, {\cal A}_{SP}(\mu_R,\mu_C) \right] 
\, ; \nonumber \\
B_{8\chi}
(\mu_C, \mu_R)&=& \left[ 1+ 
\left( \gamma^{(1)}_{88} \ln 
\left( \frac{\mu_C}{\mu_R} \right) + \Delta r_{88} 
\right) a(\mu_C)\right] \times \nonumber\\
&& \times
\left[ 1  + \frac{1}{48 \pi^2} \, 
\frac{1}{\langle 0 | \bar q q | 0 \rangle ^2(\mu_C)}
\, {\cal A}_{SP}(\mu_R,\mu_C) \right] \nonumber \\ 
&+& \frac{a(\mu_C)}{\langle 0 | \bar q q | 0 \rangle^2(\mu_C)}
\, \Delta r_{78} \, \frac{3}{16 \pi^2} 
\, {\cal A}_{LR}(\mu_R) \, .
\ea
The finite terms that appear in the
matching between the X-boson effective theory with a cut-off
and the  Standard Model regularized with the NDR scheme were calculated 
in \cite{epsprime}, 
\be
\label{rNDR}
\Delta r^{\rm NDR}_{77} = \frac{1}{8 N_c} \, ,
\quad
\Delta r^{\rm NDR}_{78} = -\frac{3}{4} \, ,
%\nonumber\\
\Delta r^{\rm NDR}_{87} = - \frac{1}{8} \, ,
\quad %&\quad&
\Delta r^{\rm NDR}_{88}   %= \frac{5}{4} C_F + \frac{3}{4N_c}  
= \frac{5}{8} N_c + \frac{1}{8N_c} \, .
\ee
The finite terms to pass from NDR to HV in the same
 basis and evanescent operators we use can be found in \cite{CFMR94}.
In the HV scheme of \cite{BJLW93,BBL96} \footnote{There is a finite
renormalization from the HV scheme of \cite{CFMR94} to the HV scheme
of \cite{BJLW93,BBL96}. 
If one uses the Wilson coefficients in the HV scheme
including the $C_F$ terms from the renormalization
of the axial current as \cite{CFMR94}, one has to add
$ - C_F$ to the diagonal terms $\Delta r_{ii}$ in (\ref{rHV}) and
$ - \beta_1 \, C_F $ to the diagonal terms in the two-loop
anomalous dimensions in (\ref{HV}).} these finite terms are
\be
\label{rHV}
\Delta r^{\rm HV-NDR}_{77} = - \frac{3}{2 N_c} \, , 
\quad %&\quad&
\Delta r^{\rm HV-NDR}_{78} = 1  \, ,
%\nonumber \\
\Delta r^{\rm HV-NDR}_{87} =  \frac{3}{2} \, , 
\quad
\Delta r^{\rm HV-NDR}_{88}  %=  C_F - \frac{1}{N_c}  
= \frac{N_c}{2}- \frac{3}{2N_c} \, .
\ee
The results for the scheme 
dependent terms  $\Delta r_{77}$ and $\Delta r_{87}$
\cite{epsprime} agree with those in \cite{DG00,KPR01}.

The $B_7$ and $B_8$ bag parameters
are  independent of $\mu$ but depend on 
$\mu_R$ and $\mu_C$, 
and these dependences only cancel in the physical value
of $\im G_E$. The $\mu_C$ dependence is artificial and 
a consequence of the normalization of the bag parameters to
the quark condensate.

At NLO in $1/N_c$ we get
\ba
\label{resbag1}
B^{NDR}_{7\chi}(\mu_C,\mu_R) 
&=& \left( 1 + \frac{a(\mu_R)}{24}  \right) 
\frac{9}{16 \pi^2} \, 
\frac{1}{\langle 0 | \bar q q | 0 \rangle^2(\mu_C)}
{\cal A}_{LR}(\mu_R)
\nonumber \\ 
&-& \frac{3}{8} \, a(\mu_C)  
\left[ 1 + \frac{1}{48 \pi^2} 
\frac{1}{\langle 0 | \bar q q | 0 \rangle ^2(\mu_C)} \, 
{\cal A}_{SP}(\mu_R,\mu_C) \right] \, ;
\nonumber \\ 
B^{NDR}_{8\chi}(\mu_C,\mu_R)
&=& \left[ 1+  \frac{1}{12} \, 
\left( 54 \ln \left( \frac{\mu_C}{\mu_R} \right) + 23  \right) 
a(\mu_C) \right]\times 
\nonumber\\
&&\times\left[ 1 + \frac{1}{48 \pi^2} 
\frac{1}{\langle 0 | \bar q q | 0 \rangle ^2(\mu_C)}
{\cal A}_{SP}(\mu_R,\mu_C) \right] 
\nonumber \\ 
&-& \, \frac{9}{64 \pi^2} \, 
 \, \frac{a(\mu_R)}{\langle 0 | \bar q q | 0 \rangle^2(\mu_C)}
{\cal A}_{LR}(\mu_R)
\ea
and in the HV scheme\cite{BJLW93,BBL96}
\ba
\label{resbag2}
B^{HV}_{7\chi}(\mu_C,\mu_R) 
&=& \left( 1 - \frac{11}{24} a(\mu_R) \right) 
\frac{9}{16 \pi^2} \, 
\frac{1}{\langle 0 | \bar q q | 0 \rangle^2(\mu_C)} \, 
{\cal A}_{LR}(\mu_R)
\nonumber \\ 
&+& \frac{33}{8} \, a(\mu_C)  
\left[ 1 + \frac{1}{48 \pi^2} 
\frac{1}{\langle 0 | \bar q q | 0 \rangle ^2(\mu_C)}
{\cal A}_{SP}(\mu_R,\mu_C) \right] \, ; 
\nonumber \\ 
B^{HV}_{8\chi}(\mu_C,\mu_R)
&=& \left[ 1+  \frac{1}{12} \, 
\left( 54  \ln \left(\frac{\mu_C}{\mu_R} \right)+  35  \right) 
a(\mu_C) \right] \,\times
\nonumber\\&&\times
 \left[ 1 + \frac{1}{48 \pi^2} 
\frac{1}{\langle 0 | \bar q q | 0 \rangle ^2(\mu_C)}
{\cal A}_{SP}(\mu_R,\mu_C) \right]  
\nonumber\\
&+& \frac{3}{64 \pi^2} \, 
\frac{a(\mu_R)}{\langle 0 | \bar q q | 0 \rangle^2(\mu_C)}
{\cal A}_{LR}(\mu_R) \, . 
\ea
We find an exact result for these B-parameters in QCD in the chiral
limit including the effects of higher dimensional
operators to all orders. The scheme dependence is also
fully taken into account. To our knowledge this
is the first time these fully model independent
expressions bag parameters are presented.
%In \cite{DG00,NAR01} only the dimension eight corrections 
%were included and the scheme dependence of the matrix elements 
%of $Q_8$ was   not included neither there nor in \cite{KPR01}.

\section{The $\Pi_{LR}^T(Q^2)$ 
Two-Point Function and Integrals over It}
\label{PILR}

There are very good data for $\Pi_{LR}^T$ \cite{ALEPH98,OPAL99} in the 
time-like region below the tau lepton mass.
They have  been extensively used previously
\cite{DG00,NAR01,DHGS98,PPR01}, see the talks
\cite{FESRtalks} for recent reviews. We consider it a good 
approximation to take this data as the chiral limit data.
Nevertheless, one can  estimate
the effect of the chiral corrections with Cauchy's
integrals around a circle of radius $4 m_\pi^2$ of the type
\ba
\oint_{4 m_\pi^2} {\rm d} s \, s^n\,  \ln^m(s) \, \Pi_{LR}^T(-s) \, 
\ea
with $n>0$, $m=0,1$.
For all the integrals we use, we have checked
that these contributions are negligible using the CHPT expressions
for $\Pi_{LR}^T(Q^2)$ at one-loop\cite{GL84}.
The discussion below is focused on the ALEPH data
but we present the OPAL results as well.

We reanalyze here the first and the second Weinberg Sum Rules 
(WSRs) \cite{WEIN67}, which are properties of QCD in the chiral limit
\cite{FSR79},
\ba
\label{WSRS}
\int_0^{\infty} {\rm d} t \,  \frac{1}{\pi}  \, \im \Pi_{LR}^T(t) =
 \int_0^{s_0} {\rm d} t \,  \frac{1}{\pi}  \, \im \Pi_{LR}^T(t) 
 + O(a^2)  = f_\pi^2 \, ; \nonumber \\
\int_0^{\infty} {\rm d} t \, t \, \frac{1}{\pi}  \, \im \Pi_{LR}^T(t)=  
  \int_0^{s_0} {\rm d} t \,  t \, \frac{1}{\pi}  \, \im \Pi_{LR}^T(t) 
 + O(a^2)
% + 2 (m_u+m_d)\langle 0 | \overline q q | 0 \rangle
 = 0 \, 
\ea
where we used the perturbative QCD result for the imaginary part
at energies larger than $s_0$, i.e. we assumed local duality above $s_0$.
 These two WSRs determine the threshold of perturbative QCD $s_0$.
We used the experimental value
for the pion decay constant  $f_\pi=(92.4 \pm 0.4)$ MeV. 

\FIGURE{
\includegraphics[angle=-90,width=0.49\textwidth]{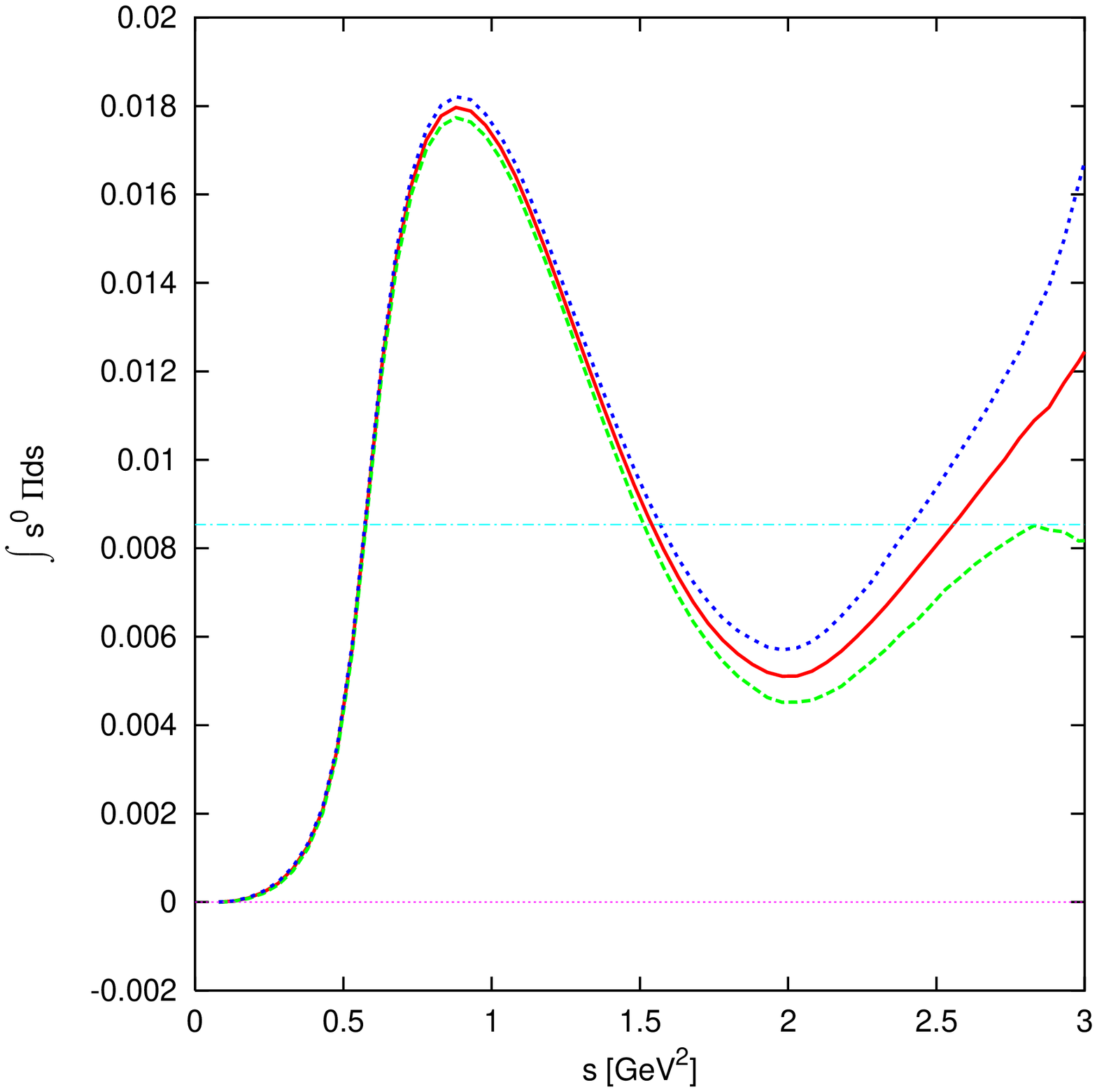}
\includegraphics[angle=-90,width=0.49\textwidth]{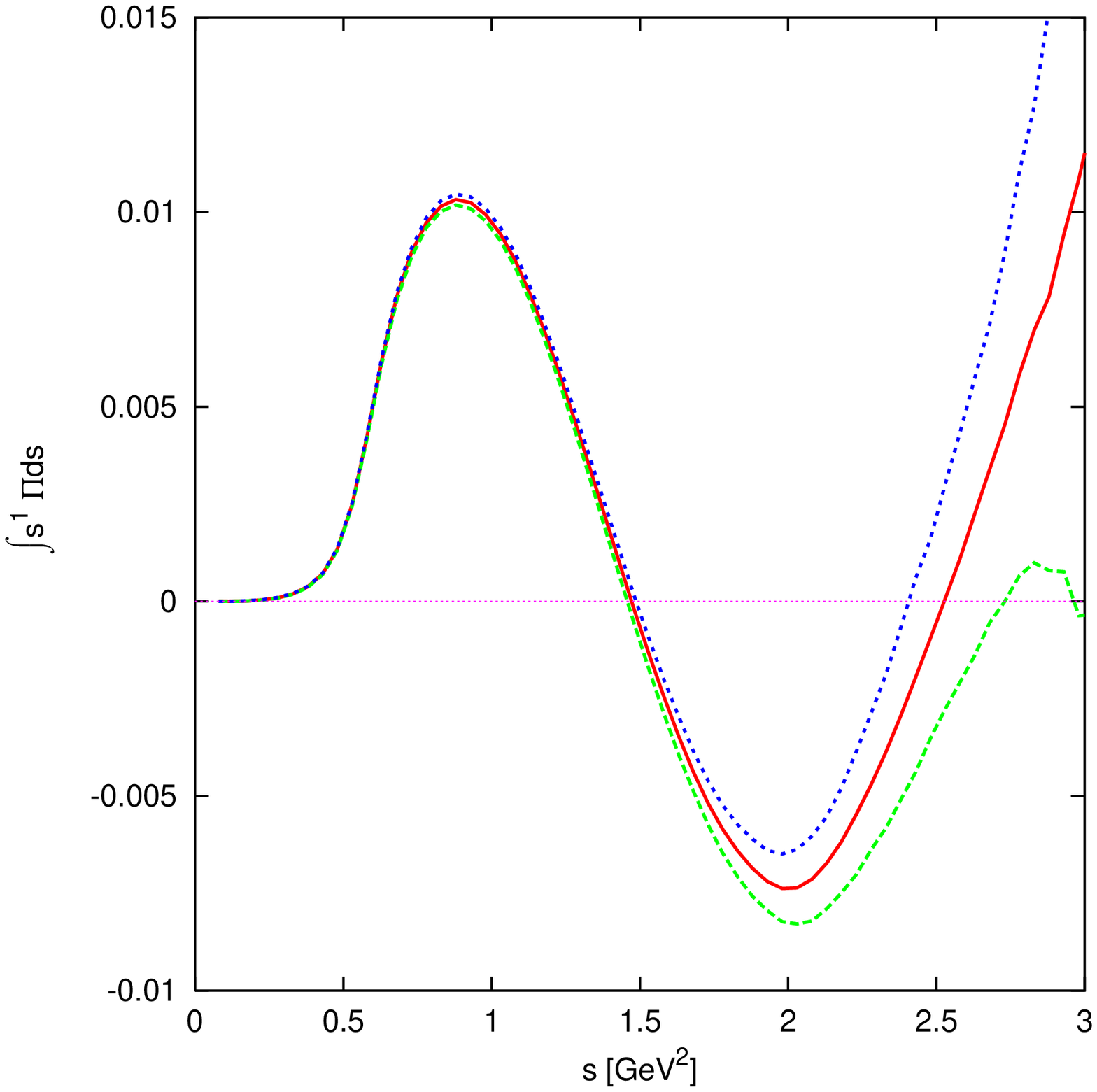}
\caption{\label{Fig1} The first and second Weinberg sum rule
as a function of the upper integration variable $s$. The central
curve corresponds to the central values of \cite{ALEPH98} while
the upper and lower curve are the one sigma errors calculated 
as described in the text.}
}%FIGURE
These two sum rules are plotted in Fig. \ref{Fig1} for the central data
values and the one sigma errors. These latter are calculated by 
generating a
distribution of spectral functions distributed according to 
the covariance
matrix of \cite{ALEPH98}. We then take the one sigma error 
to be the value
where 68\% of the distributions fall within.  All errors in the numbers
of this section and in the plots shown are
calculated in this way.

At this point, we would like to discuss where
local duality sets in: $s_0$.
As we can see from Fig. \ref{Fig1} for $s < M_\tau^2$
there are two points where (\ref{WSRS}) are satisfied,
the first one around 1.5 GeV$^2$ and the second around 2.5 GeV$^2$. 
Of course, this does not mean that  local duality
is already settled at these points as the oscillations show.
One can expect however that the violations of local duality are small
at these points. It is also obvious that local duality
will be better when the value of $s_0$ is larger.
The procedure to determine the value of $s_0$ is
repeated for each of the spectral functions generated before
and we use consistently a spectral function together
with {\em its} value of the onset of local duality.

There are several points worth making.
Though for every distribution the first duality point  
in the 1st WSR is very near the corresponding one of the 2nd WSR,
they differ by more than their error.
Numerically, when used in other sum-rules they produce results
outside the naive error.
The second duality point, $s_0\approx 2.5$ GeV$^2$
yields more stable results.
There is no a priori
reason for the value of $s_0$ to be exactly the same for different
sum rules.

Though the change from the 1st WSR
to the second is small, and even smaller if one looks at 
negative moments, when one uses large positive 
moments (the ones we need here), the deviations  
are quite sizable as we will show. This is because positive 
large moments weigh more the higher energy region
and the negative moments essentially use only information
of the low energy region.

Probably in the second duality point, local duality
has not been reached either
but certainly we should be closer to the asymptotic regime.
We therefore take the highest global duality point available,
the solution of Eq. (\ref{WSRS}), around $2.5~$GeV$^2$.
Fortunately, for the physical matrix elements, 
the additional $\log(t/\mu^2)$ in the integrand 
reduces the contribution of the data points near
the real axis for $t$ around $\mu^2$. 
This makes these sum rules much more reliable
than the single moments used in~\cite{DG00,NAR01}.

The second, and highest value with good data, value
of $s_0$ where the WSRs are satisfied
runs roughly between 2.2 GeV$^2$ and 3.0 GeV$^2$.
But not all of these values are equally probable.
If we look at the distribution of the $s_0$ values, there is a clear
peak situated around the value calculated with the central data points
but there are tails towards higher $s_0$. The widths of the peak are
essentially the same as the errors we quote.
The $s_0$ where the second WSR are mainly in the area
\ba
\label{s0}
s_0 &=& (2.53^{+0.13}_{-0.12}) \,    {\rm GeV}^2~\mbox{(ALEPH)} \,,
\quad\quad
s_0 = (2.49^{+0.17}_{-0.13}) \,    {\rm GeV}^2~\mbox{(OPAL)} \,.
\ea
and where the first WSR is satisfied in
\ba
s_0 &=& (2.56^{+0.15}_{-0.14}) \,    {\rm GeV}^2~\mbox{(ALEPH)} \,,
\quad\quad
s_0 = (2.53^{+0.17}_{-0.12}) \,    {\rm GeV}^2~\mbox{(OPAL)} \,.
\ea
These errors have been  obtained as explained above. 
In the analysis below we use all experimental
distributions with their associated value
of $s_0$ and not only those with $s_0$
in the intervals above.

The OPE of the $\Pi_{LR}^T(Q^2)$ was studied using 
the same data\cite{ALEPH98} in \cite{DHGS98}.
They obtained a quite precise determination of the
dimension six and eight higher dimensional
operators from a fit to different moments of the energy distribution.
This procedure has in principle smaller errors since one
can use the tau decay kinematic factors which suppresses the data near
the real axis but has a different local duality error.
They use $M_\tau^2$ as upper limit of the hadronic moments, 
we agree with \cite{PPR01} that one should 
use the $s_0$ where there is global duality with QCD
to eliminate possible effects of the lack of local duality 
at $M_\tau^2$. 
Another comment is that as noticed in \cite{KPR01}
the $\alpha_S^2$ corrections used in \cite{DHGS98} are
in a  different scheme \cite{LSC86}. These corrections
in the scheme used in \cite{BBL96}
are presented in the appendices.

We can determine the following higher dimensional operator
contributions (\ref{SVZLR}) 
\ba
\label{dimsix}
M_2 &\equiv& \int_0^{s_0} {\rm d} t \, t^2 \, \frac{1}{\pi}  \, 
\im \Pi_{LR}^T(t) \nonumber \\ &=& {\dis \dis \sum_{m=0}^\infty \, 
\sum_{i=1}} \langle 0| O^{(i)}_{2(m+3)}(0) | 0 \rangle(s_0) 
\, (-1)^m \, \frac{1}{2 \pi i} \, \oint_{C_{s_0}} \,{\rm d} s \, 
\frac{C^{(i)}_{2(m+3)}(s_0,-s)}{s^{1+m}} \, \nonumber \\
M_3 &\equiv& \int_0^{s_0} {\rm d} t \, t^3 \, \frac{1}{\pi}  \, 
\im \Pi_{LR}^T(t) \nonumber \\ &=& {\dis \dis \sum_{m=0}^\infty \, 
\sum_{i=1}} \langle 0| O^{(i)}_{2(m+3)}(0) | 0 \rangle(s_0) 
\, (-1)^m \, \frac{1}{2 \pi i} \, \oint_{C_{s_0}} \,{\rm d} s \, 
\frac{C^{(i)}_{2(m+3)}(s_0,-s)}{s^{m}} \, .
\ea
In Fig. \ref{fig2} we have plotted the value of $M_2$ and $M_3$
as a function of $s_0$ used in the integration, together with the one sigma
error band.
It is immediately obvious that the main uncertainty is the choice of
$s_0$ to be used. This uncertainty is increasingly important with the
increase of the moment.
\FIGURE{
\includegraphics[angle=-90,width=0.49\textwidth]{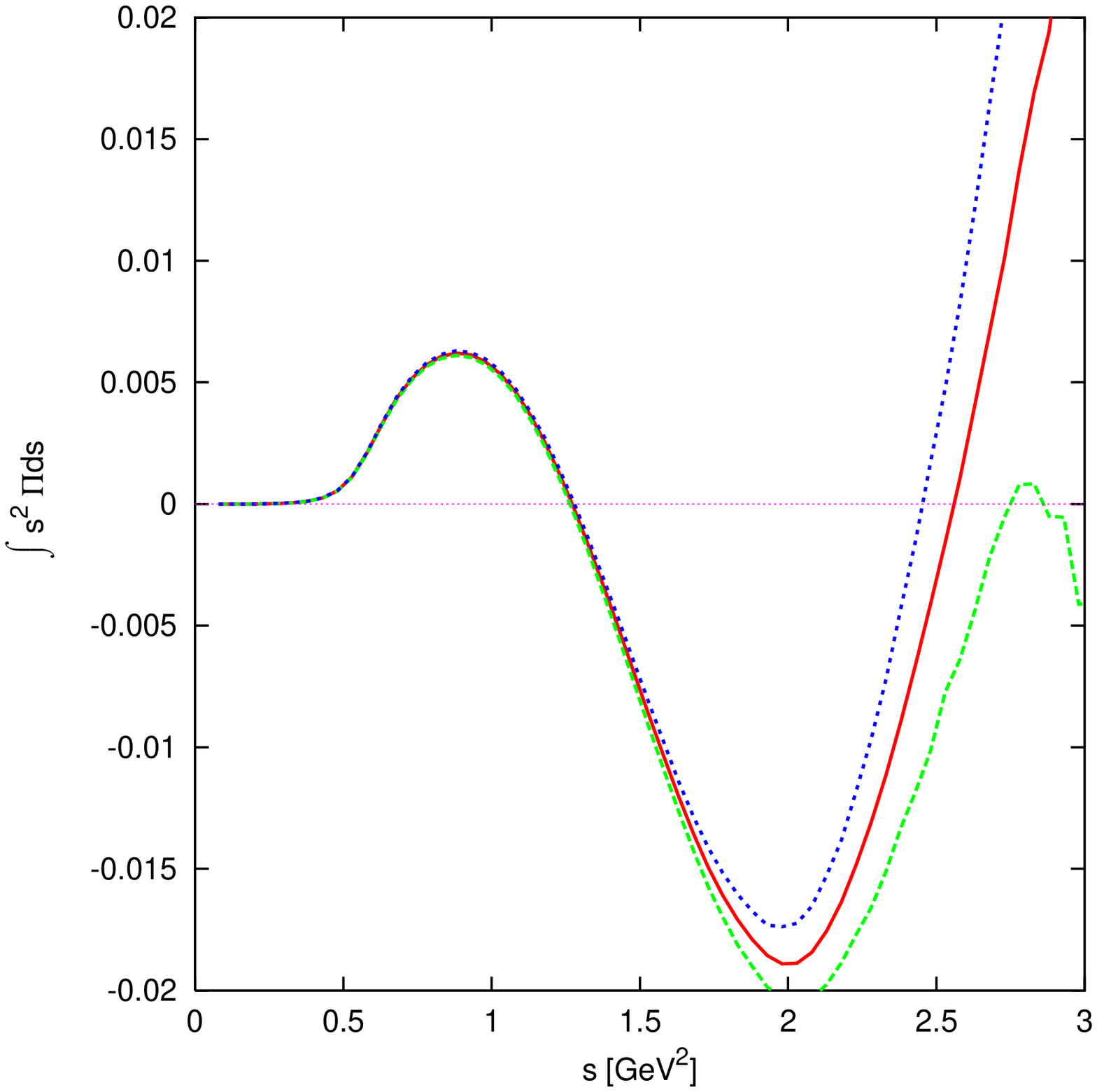}
\includegraphics[angle=-90,width=0.49\textwidth]{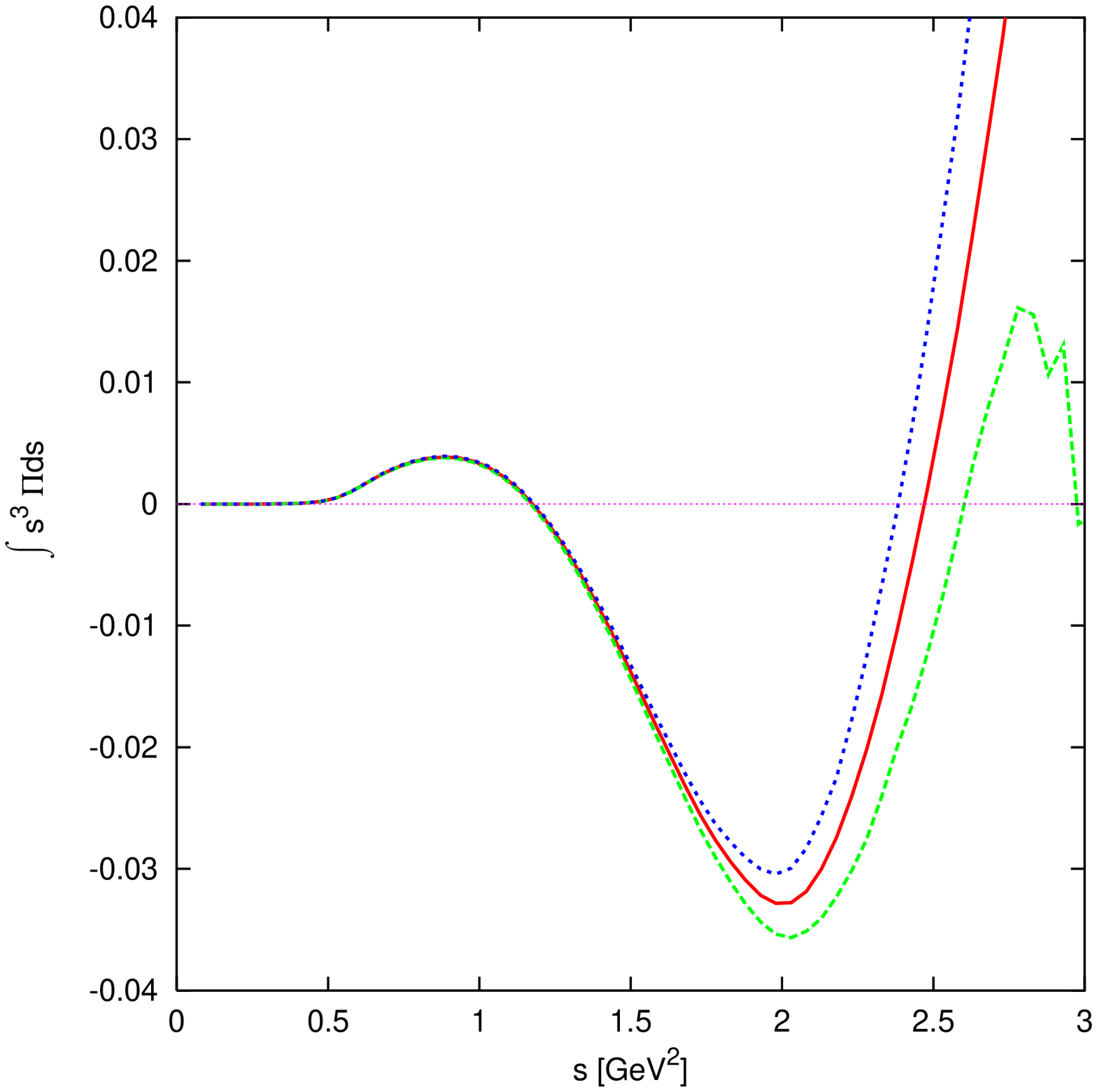}
\caption{\label{fig2} The second and third moment as a function
of the upper limit of integration $s_0$ and the one sigma variation.}
}%FIGURE

Using ALEPH data on $V-A$ spectral functions
we get for the dimension six and eight
FESR using the value for $s_0$ where the second WSR is satisfied
\ba
\label{data}
M_2 &=& -(1.7^{+1.2}_{-1.0}) \cdot 10^{-3} \, {\rm GeV}^{6} \nonumber \\
M_3 &=&  (7.2^{+5.2}_{-4.0}) \cdot 10^{-3} \, {\rm GeV}^{8} \, .
\ea
The error bars are obtained by taking 68\% of the generated distributions
within this value, only including those where the WSR can be satisfied.
The error is smaller than one would judge from Fig. \ref{fig2}
since the value of $M_2$ and $M_3$ at the value of $s_0$ where the
spectral function satisfies a WSR is much more stable than the variation
at a fixed value of $s_0$. 

Using the OPAL data we get,
\ba
\label{OPALdata}
M_2 &=& -(2.0^{+1.0}_{-1.0}) \cdot 10^{-3} \, {\rm GeV}^{6} \nonumber \\
M_3 &=& (5.2^{+4.0}_{-3.2})  \cdot 10^{-3} \, {\rm GeV}^{8} \, .
\ea

Eq. (\ref{data}) and (\ref{OPALdata})
should be compared to the results in \cite{ALEPH98,OPAL99,DHGS98}
\ba
\label{ALEPH}
M_2 &=& - (3.2 \pm 0.9) \cdot 10^{-3} \, {\rm GeV}^{6} \nonumber \\
M_3 &=& - (4.4 \pm 1.2) \cdot 10^{-3} \, {\rm GeV}^{8} \, .
\ea
The result for $M_2$ is compatible within errors 
but $M_3$ differs even in sign.
Our error bars take into account the variation of $s_0$ 
but our result for $M_3$ at the second duality point is always positive.
 
This indicates a potential problem in the determination 
of $M_3$ and higher moments (and of smaller importance in $M_2$).
As said before violations of local duality can be sizeable
for higher moments like $M_3$  even at $t\simeq M_\tau^2$
used  as upper limit of the moment integrals.
It would very helpful  to do the same  type of fit analysis done 
in \cite{DHGS98} but using the  duality point $s_0$.
Our conclusion is that moments like $M_2$, $M_3$  and higher 
are unfortunately unreliable unless one has data at higher energies. 

The integrals  which are needed for Eq. (\ref{imge}) can 
 be evaluated from the ALEPH data in the same way. We need
\ba
\label{numALEPH}
{\cal A}_{LR}(\mu_R) &\equiv& 
{\dis \int^{s_0}_0} \,
{\rm d} t \, t^2 \ln \left(\frac{t}{\mu_R^2} \right)\,  \frac{1}{\pi}
\, \im \, \Pi_{LR}^T(t) = 
(4.7_{-0.4}^{+0.5})\cdot10^{-3}~{\rm GeV}^6 \, ; 
\nonumber\\
 {\cal A}^{\rm Lower}_{LR}(\mu_R) &\equiv& 
- {\dis \int^{s_0}_0} \,
{\rm d} t \, t^2 \ln \left(1+\frac{\mu_R^2}{t} \right)\,  \frac{1}{\pi}
\, \im \, \Pi_{LR}^T(t) = 
(3.7_{-0.4}^{+0.5})\cdot10^{-3}~{\rm GeV}^6 \, ; 
\nonumber\\
 {\cal A}^{\rm Higher}_{LR}(\mu_R) &\equiv& 
{\dis \int^{s_0}_0} \,
{\rm d} t \, t^2 \ln \left(1+\frac{t}{\mu_R^2} \right)\,  \frac{1}{\pi}
\, \im \, \Pi_{LR}^T(t) =
(1.0_{-0.7}^{+0.9}) \cdot10^{-3}~{\rm GeV}^6
\ea
\FIGURE{
\includegraphics[angle=-90,width=0.49\textwidth]{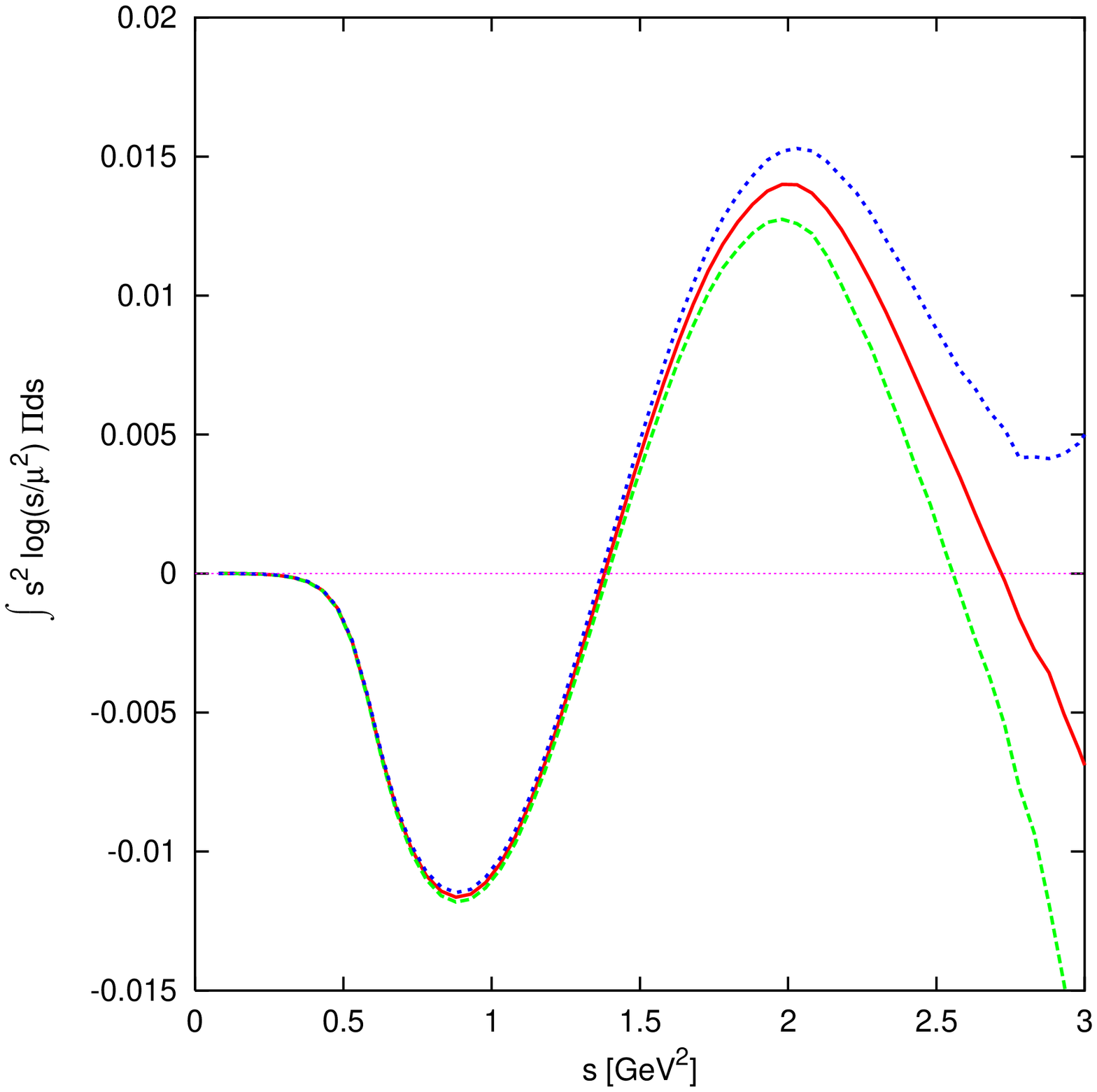}
\caption{\label{figalr} The integral over
the spectral function needed for $\im G_E$.}
}%FIGURE
at  $\mu_R$ = 2 GeV and using for each distribution its second duality
point $s_0$. Notice  the much smaller error of 
${\cal A}_{LR}$ and  ${\cal A}^{\rm Lower}_{LR}$ 
when compared with $M_2$ and $M_3$.
These values are all taken at the second duality point $s_0$ where the
second WSR is satisfied. 
We plot ${\cal A}_{LR}$
as a function of $s_0$ in Fig. \ref{figalr}.
The OPAL data give instead
\ba
{\cal A}_{LR}(\mu_R) &=& 
(4.4_{-0.3}^{+0.4})\cdot10^{-3}~{\rm GeV}^6 \, ; 
\nonumber\\
 {\cal A}^{\rm Lower}_{LR}(\mu_R) &=&
(3.8_{-0.5}^{+0.4})\cdot10^{-3}~{\rm GeV}^6 \, ; 
\nonumber\\
 {\cal A}^{\rm Higher}_{LR}(\mu_R) &=& 
(0.6_{-0.6}^{+0.8}) \cdot10^{-3}~{\rm GeV}^6
\nonumber\\
\ea

As a test, we can also calculate the electromagnetic 
pion mass difference in the chiral limit \cite{Dasetal}, 
\ba
\label{EMpion}
{\cal B}_{LR}&=&{\dis \int^{s_0}_0} \,
{\rm d} t \, t \,  \ln \left(\frac{t}{\mu_R^2} \right)\,  \frac{1}{\pi}
\, \im \, \Pi_{LR}^T(t) 
=\frac{4 \pi F_0^2}{3 \alpha_{QED}} \, 
\left(m_{\pi^0}^2- m_{\pi^+}^2\right) 
\nonumber \\ 
&=&  - (5.2 \pm 0.5) \cdot10^{-3}~{\rm GeV}^4 ~\mbox{(ALEPH)}\,; 
\nonumber\\
&=& - (5.2 \pm 0.6) \cdot10^{-3}~{\rm GeV}^4 ~\mbox{(OPAL)}\,; 
\ea
where we also used the value of $s_0$ given by the 2nd WSR.
Notice that ${\cal B}_{LR}$ does not depend on $\mu_R$
due to the second WSR (\ref{WSRS}).
The experimental number is 
\ba
\frac{4 \pi F_0^2}{3 \alpha_{QED}} \, 
\left( m_{\pi^0}^2- m_{\pi^+}^2 \right)_{\rm E.M.}
&=&
- (5.15 \pm 0.90) \cdot 10^{-3}\, {\rm GeV}^4 \, . 
\ea
where we used $F_0= (87 \pm 6)$ MeV as the chiral limit value
of the pion decay constant and 
removed the QCD contributions  \cite{ABP01}.

For comparison we quote the central values using as $s_0$ the 
second duality  point where the first WSR is satisfied
\ba
\begin{array}{rclrcl}
s_0 &=& (2.56^{+0.15}_{-0.14})~\mbox{GeV}^2\,,
&
{\cal A}_{LR}(2 {\rm GeV}) &=& (4.0_{-0.7}^{+0.6}) \cdot 10^{-3}
\, {\rm GeV}^{6} \,,\\
M_2 &=& -(0.1_{-2.8}^{+3.0}) \cdot 10^{-3}\, {\rm GeV}^{6} , 
&
 {\cal A}^{Lower}_{LR}(2 {\rm GeV}) &=& 
(2.2 \pm 2.1) \cdot 10^{-3}\, {\rm GeV}^{6} ,
\\
M_3 &=& (11_{-7}^{+9}) \cdot 10^{-3}\, {\rm GeV}^{8}\,,
&
{\cal A}^{Higher}_{LR}(2 {\rm GeV}) &=& 
 (1.8_{-1.6}^{+0.5})\cdot 10^{-3}\, {\rm GeV}^{6}.
\end{array}
\ea
for ALEPH and for OPAL
\ba
\begin{array}{rclrcl}
s_0 &=& (2.53^{+0.17}_{-0.12})~\mbox{GeV}^2\,,
&
{\cal A}_{LR}(2 {\rm GeV}) &=& (3.4_{-0.8}^{+0.7}) \cdot 10^{-3}
\, {\rm GeV}^{6} \,,\\
M_2 &=& (0.1_{-2.3}^{+2.8}) \cdot 10^{-3}\, {\rm GeV}^{6} , 
&
 {\cal A}^{Lower}_{LR}(2 {\rm GeV}) &=& 
(1.7^{+1.4}_{-1.3}) \cdot 10^{-3}\, {\rm GeV}^{6} ,
\\
M_3 &=& (10_{-6}^{+9}) \cdot 10^{-3}\, {\rm GeV}^{8}\,,
&
{\cal A}^{Higher}_{LR}(2 {\rm GeV}) &=& 
 (1.7_{-1.3}^{+1.8})\cdot 10^{-3}\, {\rm GeV}^{6}.
\end{array}
\ea

The errors are larger here. The value of $s_0$ where 
the first WSR is satisfied varies more and is somewhat larger
than the $s_0$ where the second WSR
is satisfied, this makes the last results more dependent on the
spectral function at high $t$ which have large errors.

If one tried to see the results using the first duality point, 
where less duality with QCD is expected, we get
that using the one from the 2nd WSR
\ba
\label{data1st}
\begin{array}{rclrcl}
s_0 &=& (1.47\pm 0.02)\,   {\rm GeV}^2 \;, &
{\cal A}_{LR}(2 {\rm GeV}) &=& 
(3.3 \pm 0.1) \cdot 10^{-3}\, {\rm GeV}^{6} ,
\\
M_2 &=& -(6.6\pm0.2) \cdot 10^{-3}\, {\rm GeV}^{6} , 
 &
  {\cal A}^{Lower}_{LR}(2 {\rm GeV}) 
&=&  (5.9 \pm 0.2) \cdot 10^{-3}\, {\rm GeV}^{6} ,\quad
\\ 
M_3 &=& -(12_{-2}^{+1}) \cdot 10^{-3}\, {\rm GeV}^{8} , & 
{\cal A}^{Higher}_{LR}(2 {\rm GeV})  
&=& -(2.6\pm 0.1)\, {\rm GeV}^{6}.
\end{array}
\ea
Notice that $M_2$ is not compatible with (\ref{data}) 
 with the central values
differing by more than twice the error.
The moment $M_3$ changes even sign with respect to
the second duality point showing the
problems of local duality violations for larger moments
more dramatically.
As argued before one should the largest value of $s_0$ to ensure
better local duality.
However, the physical relevant moment ${\cal A}_{LR}$ 
is much more stable  with $s_0$.

\section{The Scalar--Pseudo-Scalar Two-Point Function 
$\Pi_{SS+PP}^{(0-3)}(Q^2)$}
\label{SSPP}

In this section we discuss some of the knowledge of the spectral
function $\im \Pi_{SS+PP}^{(0-3)}(t)$ which governs
 the connected contribution to the matrix element of $Q_8$.
In the large $N_c$ limit there is no difference between the singlet and
triplet channel so the integral in (\ref{SPintegral}) is $1/N_c$ 
suppressed  and its contribution to $\im G_E$ is NNLO.
But in the scalar-pseudo-scalar sector, violations of the large $N_c$
behaviour can be larger than in the vector-axial-vector channel. 
It is therefore interesting to determine the size of this 
contribution as well.

After adding the short-distance part to the long-distance part,
the relevant integral is (\ref{imge})
\be
\label{SPintegral}
\, \frac{1}{48 \pi^2} 
\frac{1}{\langle 0 | \bar q q | 0 \rangle ^2(\mu_C)} \, 
\int_0^{\tilde s_0} 
{\rm d} t \, t \, \ln \left(\frac{t}{\mu^2} \right) 
\, \frac{1}{\pi}  \,  \im \Pi_{SS+PP}^{(0-3)}(t)  \, , 
\ee
This is the contribution
of the connected part relative to the disconnected one.
$\tilde s_0$ is the scale where in
this channel QCD duality sets in. The scale $\mu$ is the cut-off
scale. The dependence on this scale being NNLO in $1/N_c$ {\em cannot}
match the present NLO order Wilson coefficients. 

We can use models like the ones in  \cite{MOU00} 
to evaluate the scalar part of the
integrals. We only use the model there using the KLM\cite{KLM} analysis,
since only it ratifies
(\ref{weinSP2}) at a reasonable value of $\tilde s_0$.
In Fig. \ref{fig3} we plotted for that parameterization the sum rule
and the relative correction from the scalar part to the disconnected
contribution $3 \langle 0 | \overline q q | 0 \rangle^2(\mu_R)$
for $\mu=\mu_R=2$ GeV.
\FIGURE{
\includegraphics[angle=-90,width=0.49\textwidth]{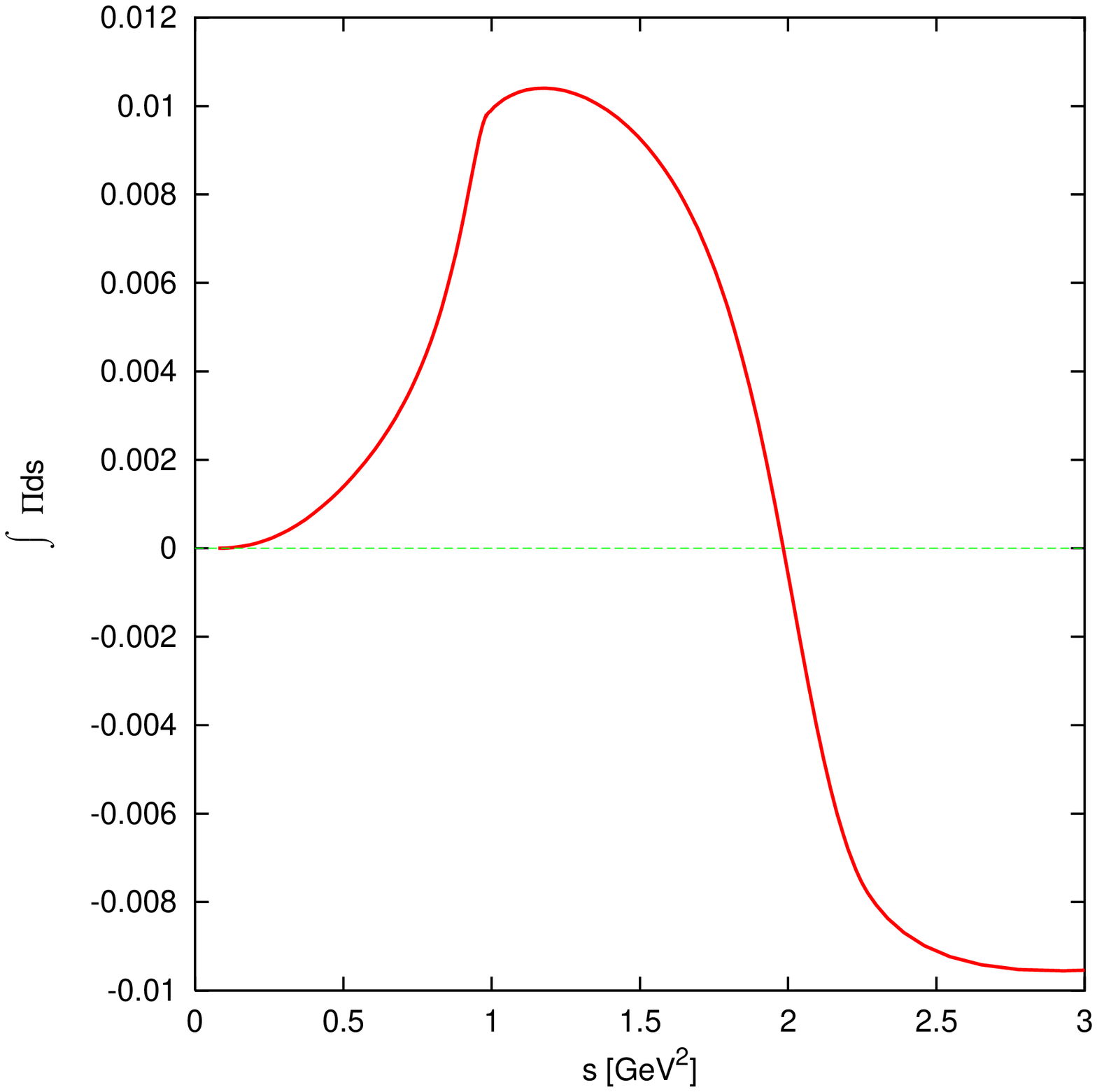}
\includegraphics[angle=-90,width=0.49\textwidth]{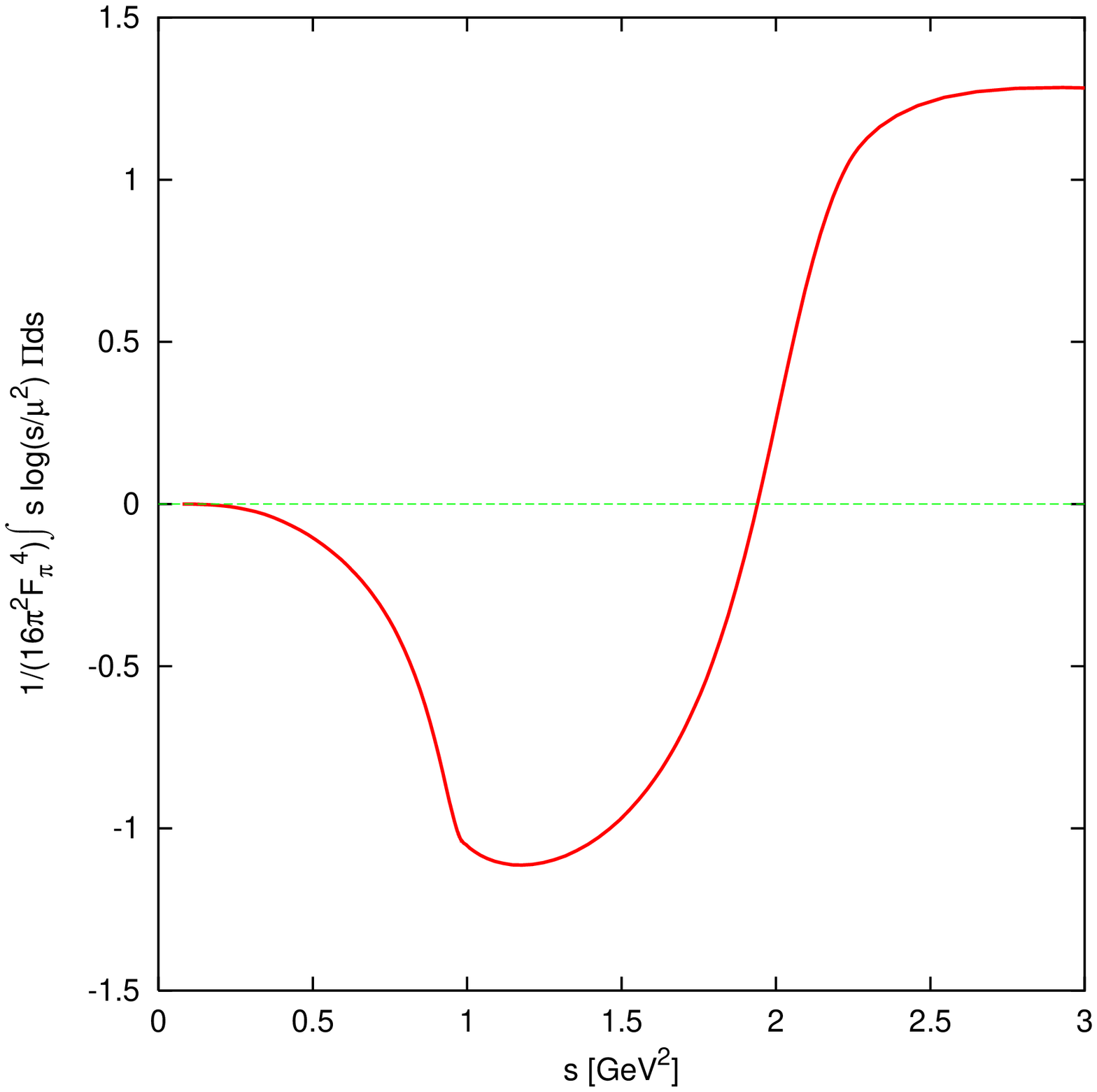}
\caption{\label{fig3} The sum rule (\ref{weinSP2})
as a function of $\tilde s_0$ for the scalar part
and the relative correction to the
disconnected contribution for $Q_8$ using the same parametrization
as a function of $\tilde s_0$ for $\mu_R=\mu=$ 2 GeV.
Notice that the correction is
small in the region where the sum rule is satisfied.}
}%FIGURE
The value of the scalar part of Eq. (\ref{SPintegral}) is about 0.18 at
$\tilde s_0 = (1.41$ GeV)$^2$.

In the large $N_c$ limit, a sensible alternative estimate 
is to use meson pole dominance.
In the pseudo-scalar sector, the U(3)$\times$U(3) symmetry
is  broken by  the chiral anomaly splitting the singlet $\eta_1$ mass
away from the zero mass for the Goldstone boson octet.

Three meson intermediate states are not studied enough to be included 
at this level, we include instead the first $\pi'$ resonance.
This means including a massless Goldstone boson 
plus the first $\pi'$ resonance
and the singlet $\eta_1$. The pseudo-scalar sum rule 
in (\ref{weinSP2}) requires the
following relation between the octet and the singlet 
couplings to the pseudo-scalar current for 
$\tilde s_0 \simeq 2.0$ GeV$^2$,
\be
F_0^2 + F_{\pi'}^2  =  F_{\eta_1}^2 \, .
\ee
Phenomenologically $F_{\pi'}^2/F_0^2<<1$ \cite{DR87}
and $F_0 \simeq F_{\eta_1}$.

We can introduce a scalar meson octet $S_8$ and a singlet $S_1$
using the methods of \cite{Eckeretal}.
The coupling constant for the octet can be denoted by 
$c_m$ and has been
estimated in \cite{Eckeretal,ENJL} to be about $(43\pm 14)$ MeV.
In fact, the sum rule (\ref{weinSP1}) is a property of QCD
and relates in this approximation $c_m$ to $F_0$
\be
c_m^2 = \frac{1}{8} \left[ F_0^2 + F_{\pi'}^2 + \cdots \right]
= \frac{F_{\eta_1}^2}{8} \simeq \frac{F_0^2}{8} 
\ee
which numerically agrees quite well with the phenomenological 
estimate.

The scalar sum rule in (\ref{weinSP2}) requires 
the singlet and the octet 
components to have the same coupling leading to
a relative correction from the scalar integral to the disconnected
contribution of
\be
\frac{1}{12\pi^2} \frac{F_{\eta_1}^2}{F_0^4} \, 
\left[M_{S_1}^2 \ln\left(\frac{M_{S_1}}{\mu}\right)
-M_{S_8}^2\ln\left(\frac{M_{S_8}}{\mu}\right)\right].
\ee
using both sum rules and the lowest meson dominance approximation.

The contribution from the pseudo-scalar 
connected two-point function  relative the disconnected
contribution can then be evaluated to
\ba
\frac{1}{12\pi^2} \frac{F_{\eta_1}^2}{F_0^4} \,
\left[ 
 M_{\eta_1}^2 \, \ln\left(\frac{M_{\eta_1}}{\mu}\right)\, 
-\frac{F_{\pi'}^2 }{F_{\eta_1}^2} \, M_{\pi'}^2
\ln\left(\frac{M_{\pi'}}{\mu}\right) 
\right] \, .
\ea
The contribution of the $\pi'$ is negligible.

As said before the scale $\mu$ is free and cannot
be at present matched with OPE QCD since it is a NNLO order in $1/N_c$
effect.
The scale independence is  reached when the sum rule
\ba
\label{MTILDE1}
\tilde M_1 &=& 
M_{S_8}^2-M_{S_1}^2-M_{\eta_1}^2+\frac{F_{\pi'}^2}{F_{\eta_1}^2}
M_{\pi'}^2 =0
\ea
which is $O(N_c^2 \alpha^2)$ (\ref{M1sumrule}) is fulfilled. 
This sum rule is very well satisfied in the linear $\sigma$ model,
see e.g. \cite{MOU00}. 
%Being $O(\alpha^2)$, it is also
%well satisfied for the values of $Q_7$ and $Q_8$ operators
%in the literature \cite{DG00,NAR01,KPR01,CDGM01,DGGM99}.

The masses $M_{\eta_1}\simeq 0.86$ GeV (chiral limit value)
$M_{\pi'}\simeq 1.3$ GeV are known.
The masses of the singlet and octet of scalars are not so well known.
Using  $M_{S_1} = M_{\sigma}\simeq 0.5$ GeV and 
$M_{S_8} = M_{a_0(980)}
\simeq 0.98$ GeV the correction to the disconnected contribution
is almost independent of the scale $\mu$, neglecting the $\pi^\prime$
and is independent of $\mu$ for $F_{\pi^\prime}^2/F_0^2 = 0.017$.
The relative scalar contribution is about $0.40$
and the relative pseudoscalar is $-0.71$
with a total relative contribution of about $-0.31$.
The scalar contribution is within errors compatible with 
the earlier estimate.

{}From the discussion here it can be seen that there is 
a possibly sizable correction but we expect it to be 
smaller than $-40$\%.  

Of course, the scale dependence
left in (\ref{SPintegral}) is unsatisfactory in principle 
but small since the sum rule (\ref{MTILDE1}) is quite well satisfied.

\section{Numerical Results for the Matrix-Elements and Bag Parameters}

The vacuum expectation value in the chiral limit of $Q_7$ itself
is related directly to $B_{7 \chi}$.  This allows us 
to obtain\footnote{The
analytical formulas  are  in agreement with \cite{DG00,NAR01,KPR01}
for the  scheme dependent terms in $Q_7$ matrix elements but 
not for the $Q_8$ ones in \cite{DG00,NAR01} and they were 
not included  in \cite{KPR01}.}
\ba
\label{numresult}
\langle 0 | O_6^{(1)}(0) | 0 \rangle^{NDR} (\mu_R) &=&
- \left(1+\frac{1}{24}  a(\mu_R) \right) \, \frac{3}{32\pi^2} 
{\cal A}_{LR}(\mu_R) \nonumber \\
&+& \frac{1}{48}\, a(\mu_R) \, 
\left[ 3  \langle 0 | \overline q q | 0 \rangle^2  (\mu_R)
+ \frac{1}{16 \pi^2} 
{\cal A}_{SP}(\mu_R,\mu_C) \right] \, ;
\ea
\ba
\label{numresultHV}
\langle 0 | O_6^{(1)}(0) | 0 \rangle^{HV} (\mu_R) &=&
-\left(1- \frac{11}{24} a(\mu_R)  \right)
\frac{3}{32\pi^2} \, {\cal A}_{LR}(\mu_R) \nonumber \\
&-& \frac{11}{48}\, a(\mu_R) \, 
\left[ 3 \langle 0 | \overline q q | 0 \rangle^2  (\mu_R) \, 
 + \frac{1}{16 \pi^2} 
{\cal A}_{SP}(\mu_R,\mu_C) \right] \, .
\ea

For the numerics, 
we use the value of the condensate obtained in the $\overline{MS}$
scheme in \cite{BPR95}, 
\ba
\label{factor6}
\langle 0 | \overline q q | 0 \rangle (2 {\rm GeV})&=&
- (0.018\pm 0.004) \, {\rm GeV}^3 \,,
\ea
the numerical results of Eq. (\ref{numALEPH}),
\be
a(2~GeV) = 0.102
\ee
and neglect, in first approximation, 
the integral over the scalar--pseudo-scalar two-point function.

The weighted average of the first and second WSR results
for ${\cal A}_{LR}(2 {\rm GeV})$ from ALEPH data is
\be
{\cal A}^{\rm ALEPH}_{LR}(2{\rm GeV})
= (4.5\pm0.5) \cdot10^{-3}~{\rm GeV}^6  
\ee
 and from OPAL data
\be
{\cal A}^{\rm OPAL}_{LR}(2{\rm GeV})
= (4.2\pm0.4) \cdot10^{-3}~{\rm GeV}^6  \, . 
\ee
Though the systematic errors aver very correlated, sicne the
central values are very similar we take the simple average of
 both results as our result 
\be
{\cal A}_{LR}(2{\rm GeV})
= (4.35\pm0.50) \cdot10^{-3}~{\rm GeV}^6  \,  
\ee
and obtain
\ba
\label{numO1}
\lefteqn{\langle 0 | O_6^{(1)}(0) | 0 \rangle^{NDR} ( 2 {\rm GeV}) =
-(4.0\pm0.5)\cdot10^{-5}~{\rm GeV}^6}&&
\nonumber\\
 &=&
\left(-(4.2\pm0.5)+(0.2\pm0.1)\right)
\cdot10^{-5}~{\rm GeV}^6 
\nonumber\\
&=&
\left((-3.3\pm0.5)+(-0.9\pm0.8)+(0.2\pm0.1)\right)
\cdot10^{-5}~{\rm GeV}^6
\ea
and
\ba
\label{numO2}
\lefteqn{\langle 0 | O_6^{(1)}(0) | 0 \rangle^{HV} ( 2 {\rm GeV}) =
-(6.2\pm1.0)\cdot10^{-5}~{\rm GeV}^6}&&
\nonumber\\
 &=&
\left((-3.9\pm0.5)-(2.3\pm0.9)\right)
\cdot10^{-5}~{\rm GeV}^6 
\nonumber\\
&=&
\left((-3.1\pm0.5)+(-0.8\pm0.7)-(2.3\pm0.9)\right)
\cdot10^{-5}~{\rm GeV}^6
\ea
where we quote, namely, the total result, the integral and the vacuum 
expectation value
separately and in the last case also the long and 
short-distance part of the  integral separately. 

The short-distance part of the integral,
the second term in the above, is the contribution of
all higher dimensional operators. We find that its contribution
is between a few \% up to 35 \% depending on the value of $\mu$.
At $\mu=2$ GeV it is somewhat larger than the error
on the integral cut-off at $\mu$.

Similarly, the matrix-element of $Q_8$ is directly related
to $B_8$ and we obtain\footnote{We disagree in this case with the results
in \cite{DG00,NAR01,KPR01} because of the scheme dependent
terms. These references also disagree with each other.}
\ba
\label{O6operator}
 \langle 0 |  O_6^{(2)}(0) | 0 \rangle^{NDR} ( 2 {\rm GeV} ) 
&=& \left[ 1+ \frac{23}{12} \, a(2 {\rm GeV}) \right] \times
\nonumber\\ && \times 
\left[ 3 \, \langle 0 | \overline q q | 0 \rangle^2  (2 {\rm GeV})
+ \frac{1}{16 \pi^2} {\cal A}_{SP}(2 \rm{GeV},2\rm{GeV}) \right]\, 
\nonumber \\ 
&-&   \frac{27}{64\pi^2} \, a(2 {\rm GeV})  \, 
{\cal A}_{LR}( 2 {\rm GeV}) \, .
\ea
\ba
 \langle 0 |  O_6^{(2)}(0) | 0 \rangle^{HV} ( 2 {\rm GeV} ) 
&=& 
\left[ 1+ \frac{35}{12} \, a(2 {\rm GeV}) \right] \times
\nonumber\\&&\times
\left[ 3 \, \langle 0 | \overline q q | 0 \rangle^2  (2 {\rm GeV})
+ \frac{1}{16 \pi^2} {\cal A}_{SP}(2 \rm{GeV},2\rm{GeV}) \right]\, 
\nonumber \\  &+&   
\frac{9}{64\pi^2} \, a(2 {\rm GeV})  \,  {\cal A}_{LR}( 2 {\rm GeV})
\,  .
\ea

Using the same input as above we obtain
\ba
 \langle 0 |  O_6^{(2)}(0) | 0 \rangle^{NDR} ( 2 {\rm GeV} ) 
&=& (1.2 \pm 0.5 ) \cdot 10^{-3} \, {\rm GeV}^6 \,, \nonumber \\
 \langle 0 |  O_6^{(2)}(0) | 0 \rangle^{HV} ( 2 {\rm GeV} ) 
&=& (1.3 \pm 0.6 ) \cdot 10^{-3} \, {\rm GeV}^6 \,, 
\ea
where the contribution of the integral over $\im \Pi_{LR}^T$
is at the 1\% level and thus totally negligible.

Another combination of these two matrix-elements can also
be obtained from an integral over the ALEPH data\cite{BJLW93,BBL96}
by  putting  (\ref{D6_2}) and (\ref{SVZLR}) in (\ref{dimsix}) 
including also the $\alpha_S$ correction of the appendix
\footnote{We thank Vincenzo Cirigliano, John Donoghue,
Gene Golowich, Marc Knecht, Kim Maltman, Santi Peris, and
Eduardo de Rafael for pointing out an error in the matching 
coefficients in the previous version of our paper.
Our result agrees with the result found in \cite{CDGM01}}:
\ba
\label{sumrule6}
M_2 &=&\int_0^{s_0} {\rm d} t \, t^2 \, \frac{1}{\pi}  \, 
\im \Pi_{LR}^T(t) = {\dis \sum_{i=1}} \, C^{(i)}_{6}(s_0,s_0) \, 
O^{(i)}_6(s_0) \nonumber \\ 
&=& - \frac{4 \pi^2}{3} 
\, a(s_0) \left[ 2 \left(1 +  \frac{13}{8} \, a(s_0) \right) \, 
\langle 0 |  O_6^{(1)}(0) | 0 \rangle^{NDR} (s_0) 
\right. \nonumber \\
&+& \left.  \, \left(1+ \frac{25}{8} a(s_0) \right) 
\, \langle 0 | O_6^{(2)}(0) | 0 \rangle^{NDR} (s_0) \right] \, 
\nonumber \\ 
&=& - \frac{4 \pi^2}{3} 
\, a(s_0) \left[ 2 \left(1 +  \frac{41}{8} \, a(s_0) \right) \, 
\langle 0 |  O_6^{(1)}(0) | 0 \rangle^{HV} (s_0) 
\right. \nonumber \\
&+& \left.  \, \left(1+ \frac{21}{8} a(s_0) \right) 
\, \langle 0 | O_6^{(2)}(0) | 0 \rangle^{HV} (s_0) \right] \, .
\nonumber \\
&=& - 4 \pi^2 a(s_0) \left[ \left(1+\frac{61}{12} a(s_0) \right)
\, \left[ \langle 0 | \overline q q | 0 \rangle^2  (s_0)
+ \frac{1}{48 \pi^2} {\cal A}_{SP}(s_0,s_0) \right]\,\right.
\nonumber \\
&-& \left. \left( 1 +\frac{47}{12} a(s_0) \right) \frac{1}{16 \pi^2}
{\cal A}_{LR}(s_0)\right]
\ea
The right hand-side is physical and we cheked that
is independent of the scale $s_0$ and scheme. 
We can therefore evaluate it at $s_0=4$~GeV$^2$.
The contribution from 
$\langle 0 |  O_6^{(1)}(0) | 0 \rangle^{NDR} (s_0)$
is numerically very small and we obtain 
\ba
M_2 = -(2.0\pm 0.9)\cdot10^{-3}~\mbox{GeV}^6\,, 
\ea
perfectly compatible within errors both 
with the result obtained from
the data in Eq. (\ref{data}) and
with the result (\ref{ALEPH}). This confirms our
results on the size of the 
integral over $\im \Pi_{SS+PP}(t)$, which can therefore
be considered negligible within the present accuracy
of the disconnected contribution and $M_2$.

There is another sum rule which combines the two
matrix elements, 
\ba
\tilde M_1 &=& 
\int_0^{\tilde s_0} {\rm d} t \, t \, \frac{1}{\pi}  \, 
\im \Pi_{SS+PP}(t) = - 
{\dis \sum_{i=1}} \, \tilde C^{(i)}_{6}(s_0,s_0) \, 
  O^{(i)}_6(s_0) \nonumber \\ 
&=& - \frac{\pi^2}{4} \, a(s_0)^2 \, \left[ 
211 \, \langle 0 |  O_6^{(2)}(0) | 0 \rangle  (\tilde s_0) 
+  90 \, 
\langle 0 |  O_6^{(1)}(0) | 0 \rangle  (\tilde s_0) \right] 
+ O(a^3) \, . \nonumber \\
\ea
For the calculation of the coefficients see Appendix \ref{AppB}.
This sum rule is much less accurate than $M_2$
since the leading terms are  $\alpha_S^2$
and the value of $\tilde M_1$ is not known directly either. 
Therefore we don't use it.

The numerical estimates of the disconnected part, ${\cal A}_{SP}$,
given above change these numbers somewhat but within the 
errors quoted.

These results can also be expressed in terms of the bag parameters:
\ba
\label{numbag}
B_{7 \chi}^{NDR}(2 {\rm GeV})&=& 0.75 \pm 0.20 \; ;\quad
B_{7 \chi}^{HV}(2 {\rm GeV})= 1.15 \pm 0.30\;  \nonumber \\
B_{8 \chi}^{NDR}(2 {\rm GeV})&=& 1.2 \pm 0.3  \; ; \quad
B_{8 \chi}^{HV}(2 {\rm GeV})= 1.3 \pm 0.4\,.
\ea
We can also express it in terms of $\im G_E$:
\ba
F_0^6~\im G_E &=& \im\tau (-2.1\pm0.9)~10^{-6}~\mbox{GeV}^6\,
\ea
which is quite compatible with the estimate in \cite{epsprime}.
 
\section{Comparison with earlier results}
\label{comparison}

To compare with other results
in the literature we propose to use the VEVs
$\langle 0| O_6^{1)} | 0 \rangle$ and
$\langle 0| O_6^{(2)} | 0 \rangle$. The reason is that
these quantities are what \cite{DG00,NAR01,KPR01} and we directly 
compute.  The matrix elements of $K\to \pi\pi$ through
$Q_7$ and $Q_8$, in the chiral limit,\footnote{See \cite{epsprime} for
the definition of $M_2[Q_7]$ and $M_2[Q_8]$.} can be expressed as follows
\cite{epsprime}
\ba
M_2[Q_7](\mu_R)&=&
\langle (\pi\pi)_{I=2}| Q_{7} | K^0 \rangle (\mu_R) =
-\sqrt{\frac{2 }{3}} \, 
\frac{\langle 0|\overline q q | 0 \rangle^2(\mu_C)}{F_0^3} \, 
\, B_{7\chi}(\mu_C,\mu_R)  \, \nonumber  \\
&=&  2 \sqrt 6 \,\frac{ \langle 0| O_6^{(1)} | 0 \rangle_\chi (\mu_R)}
{F_0^3}
\, ; \nonumber \\
M_2[Q_8](\mu_R)&=&
\langle (\pi\pi)_{I=2}| Q_{8} | K^0 \rangle (\mu_R) = 
-\sqrt 6 \, \frac{\langle 0|\overline q q | 0 \rangle^2(\mu_C)}{F_0^3} \, 
\, B_{8\chi}(\mu_C,\mu_R)  \, \nonumber  \\
&=& - \frac{\sqrt 6}{3} \, \frac{\langle 0| O_6^{(2)} | 0 \rangle_\chi 
(\mu_R)}{F_0^3} \, .
\ea

The lattice results\cite{DGGM99} are from computed $K \to \pi$ matrix elements
and use the physical values of $f_K$ and $f_\pi$
to convert into $K\to \pi \pi$. 
Since we and  \cite{DG00,NAR01,KPR01} compute
in the chiral limit, this amounts to a large 
spurious factor $f_K f_\pi^2/F_0^3 \simeq 1.6$ 
of difference when comparing $K \to \pi \pi$ matrix elements
or  $f_K f_\pi /F_0^2 \simeq 1.4$ when comparing
$K \to  \pi $ matrix elements.
Usually this factor is not taken into account.
Moreover each group uses different conventions,
either the chiral limit value of $f_{\pi}$ and $f_K$ or their
physical value. 
We give the lattice results for $K\to\pi$ rescaling with the factor 
above.
\TABLE{
%\begin{center}
\begin{tabular}{|c|c|c|}
\hline
Reference&$- 10^{5} \langle 0| O_6^{(1)}|0 \rangle^{NDR}_\chi
\, {\rm GeV}^{-6}$&
$10^{3}\langle 0| O_6^{(2)}|0 \rangle^{NDR}_\chi
\, {\rm GeV}^{-6}$ 
 \\ 
\hline
$B_7=B_8=1$                      &$5.4\pm2.2$& $1.0 \pm 0.4$   \\
This work (SS+PP=0)              &$4.0\pm0.5$& $1.2  \pm 0.5$  \\  
Knecht et al.  \cite{KPR01}      &$1.9\pm0.6$& $3.5 \pm 1.1$   \\   
Cirigliano et al. \cite{CDGM01}  &$2.7\pm1.7$& $2.2 \pm 0.7$   \\  
Donoghue et al.\cite{DG00}       &$4.3\pm0.9$& $1.5 \pm 0.4$      \\  
Narison        \cite{NAR01}      &$3.5\pm1.0$& $1.5 \pm 0.3$      \\  
lattice        \cite{DGGM99}     &$2.6\pm0.7$& $0.74 \pm 0.15$    \\
ENJL  \cite{epsprime}            &$4.3\pm0.5$& $1.3\pm0.2$\\
\hline
\end{tabular}
%\end{center}
\caption{The values of the VEVs in the NDR scheme at $\mu_R=2$ GeV.}
}%TABLE
\TABLE{
%\begin{center}
\begin{tabular}{|c|c|c|}
\hline
Reference&$- 10^{5} \langle 0| O_6^{(1)}|0 \rangle^{HV}_\chi
\, {\rm GeV}^{-6}$&
$10^{3}\langle 0| O_6^{(2)}|0 \rangle^{HV}_\chi
\, {\rm GeV}^{-6}$ 
 \\ 
\hline
$B_7=B_8=1$                      &$5.4\pm2.2$ & $1.0 \pm 0.4$    \\
This work (SS+PP=0)              &$6.2\pm 1.0$& $1.3 \pm 0.6$    \\  
Knecht et al.  \cite{KPR01}      &$11.0\pm2.0$& $3.5 \pm 1.1$      \\   
Cirigilano et al.\cite{CDGM01}   &$8.2\pm0.9$& $2.4\pm0.7$      \\  
lattice        \cite{DGGM99}     &$4.3\pm 1.1$& $0.8 \pm 0.2$    \\
ENJL \cite{epsprime}             &$7.1\pm0.9$ & $1.4\pm0.2$\\
\hline
\end{tabular}
%\end{center}
\caption{The values of the VEVs in the HV scheme at $\mu_R=2$ GeV.}
}%TABLE

We agree with the old results by \cite{DG00,CDG00}
including the higher order operators. In fact, they also make 
an estimate of their contribution which agrees with our full
calculation. We also agree reasonably well with \cite{NAR01}.
We agree borderline with their new 
results\cite{CDGM01} within errors, though their
central value is almost twice ours for $Q_8$.

We do not agree with \cite{KPR01} even within errors.
But if the $O(a^2)$ correction in (\ref{sumrule6}) 
is taken into account, their result for $O_6^{(2)}$
goes to $(2.7\pm 0.9) \, 10^{-3}$ and we are borderline also
within errors though the  central value is more than twice ours.

We find a systematic a factor around
1.5  to 1.8 compared lattice results\cite{DGGM99} for both matrix elements.
They are compatible when we take the errors on both 
the lattice and our results into account and the
fact that the lattice numbers are not
in the chiral limit. The corrections for the latter
we included only partially
with the physical values of
$f_K$ and $f_\pi$.

We have not quoted the results from the CHPT large $N_c$ approach\cite{HKPSB98}
and the chiral quark model \cite{Trieste} because they are away from the
chiral limit.

Our results are very compatible with our earlier work.
In \cite{epsprime} we used the ENJL model and a very low value of $\mu$
to estimate the same matrix-elements.
The underlying reason for the agreement is that the operators $Q_7$ and $Q_8$
mix quite strongly and the values of the matrix-element of $Q_8$
at the low scale $\mu\approx 0.8$ GeV
dominate the values of the matrix-elements
of $Q_7$ and $Q_8$ at the higher scale $\mu=2$~GeV.
The matrix element of $Q_8$ obtained here is the same
as the one we used in \cite{epsprime}, since the effect
of ${\cal A}_{SP}$ is estimated to be moderate here.

\subsection{R\^ole of Higher Dimensional Operators}
\label{higherdim}

We clarified here the role of the higher than six dimensional operators,
an issue raised in \cite{CDG00}. In our scheme they remove
the $\mu$-dependence which is not covered by the renormalization group.

The effect of higher dimension operators in our approach is to add 
${\cal A}_{LR}^{\rm Higher}(\mu)$
to the low energy contribution ${\cal A}_{LR}^{\rm Lower} (\mu)$,
these are defined in Eq. (\ref{numALEPH}),
\ba
{\cal A}_{LR}(\mu) &=&
{\cal A}_{LR}^{\rm Higher}+{\cal A}_{LR}^{\rm Lower}
= \int_0^{s_0} {\rm d} t \, t^2 \, \ln \left(\frac{t}{\mu^2} \right) 
\, \frac{1}{\pi}  \, \im \Pi_{LR}^T(t) \, 
\ea
where $\mu$ is an Euclidean cut-off. 
It is clear than the contribution of higher than 
dimension six operators
is less important only for values of $\mu^2$ larger than $s_0$,
where $\im \Pi^T_{LR}$ vanishes because of local duality.
 In Figure \ref{fig4} we plot the two separate contributions
and the sum 
as a function of $\mu$.
\FIGURE{
\includegraphics[angle=-90,width=0.49\textwidth]{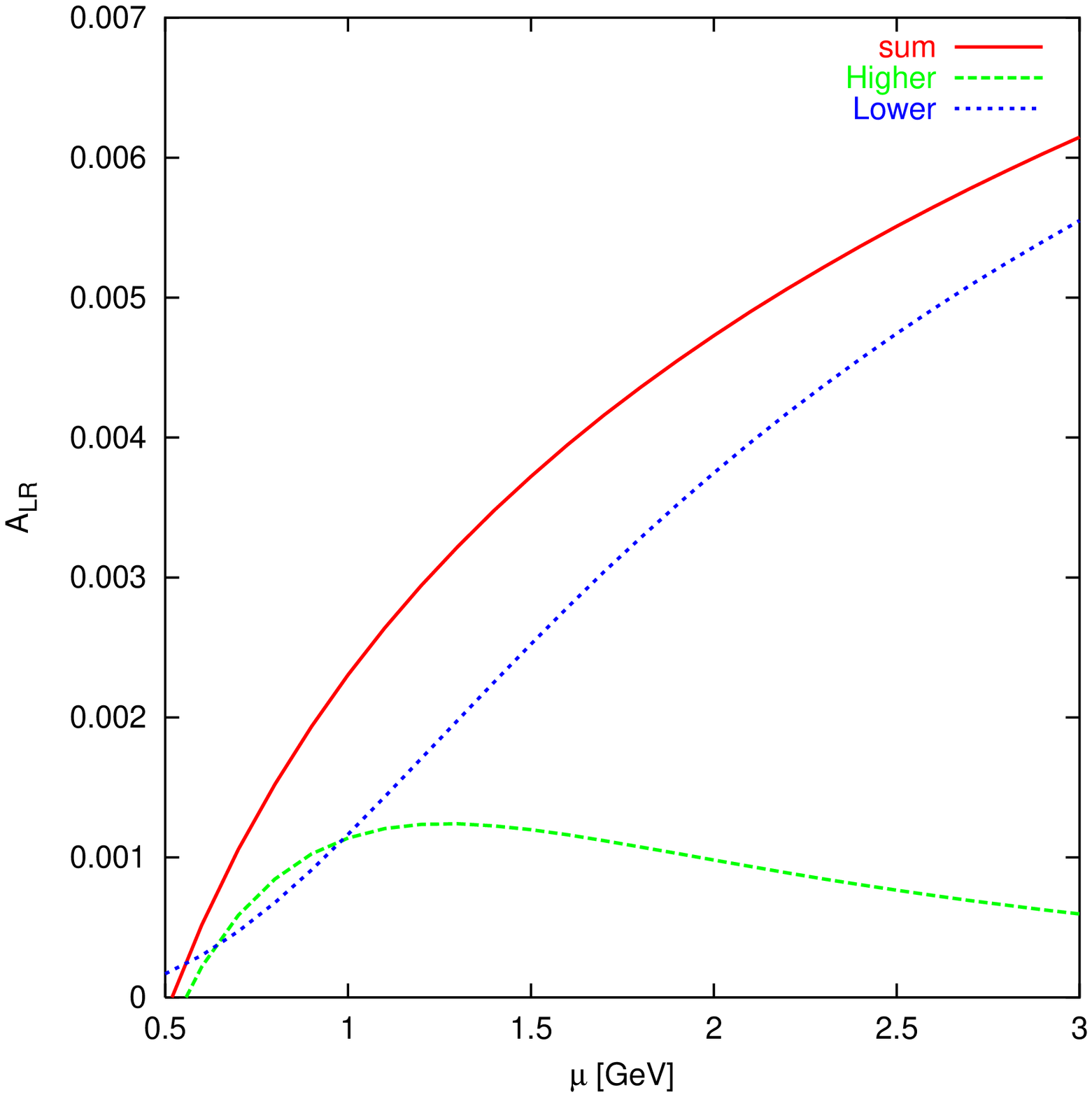}
\caption{\label{fig4} The separate contributions
to ${\cal A}_{LR}$ and the sum. ``Higher'' labels the effect of
the higher than six dimensional operators in the short-distance
contribution and ``Lower'' the long-distance part.}
}%FIGURE
{}From the figure
we can see for $\mu$ larger than 2 GeV the contribution
of {\em all} higher dimensional operators is less than 25 \%.
We agree with \cite{CDG00} that for the matrix elements that involve
integrals of $\im \Pi^T_{LR}$ one has to go to
such values of $\mu$ to disregard the contribution of 
higher dimensional operators. The contribution we find is somewhat smaller
than in \cite{CDG00} since we include the effect of {\em all}
higher order operators, not just dimension eight.

The high value of $\mu$ is set by the threshold of
perturbative QCD $s_0$ which depends very much on the 
spectral function and on the integrand behaviour.
In fact, from \cite{PR00}
one can  see that  relevant spectral function for the 27-plet coupling
reaches the perturbative QCD behaviour very soon, from 0.7 GeV
to 1 GeV.  The  OPE 
matched impressively well with the hadronic ansatz at such low values
with just dimension six operators.
Therefore  though higher dimensional operators appear
one  can expect smaller contributions
in cases like $G_{27}$ and $ \re G_8$.

The matrix-elements studied in this paper might 
be special in the sense
that they follow from integrals over spectral functions which
have no contributions at short-distances from the unit operator
or the dimension four operators. As the good
matching at low scales in the example in \cite{PR00}
shows, the other quantities which have these contributions might
have much smaller higher dimension effects.

\subsection{The Large  $N_c$ Limit}

In the  large $N_c$ limit, the spectral functions have only poles:
\ba
\label{LRNC}
 \frac{1}{\pi} \, \im \Pi^T_{LR}(t)&=&
-F_0^2 \delta(t) - \sum_A^\infty f_A^2 M_A^2 \delta(t-M_A^2) 
+ \sum_V^\infty f_V^2 M_V^2 \delta(t-M_V^2) 
\ea
and
\ba
\label{SPNC}
 \frac{1}{\pi} \, \im \Pi^{(0-3)}_{SS+PP}(t)&=&
4 \frac{\langle 0 | \overline q q | 0 \rangle^2 (\mu_C)}{F_0^4} \,
\left[ 8 \sum_{S_1}^\infty \tilde c_{m,S_1}^2 \delta(t-M_{S_1}^2) 
- 8 \sum_{S_8}^\infty c_{m,S_8}^2 \delta(t-M_{S_8}^2) 
\right. \nonumber \\
&-& \left. \sum_{\eta_1}^\infty F_{\eta_1}^2 \delta(t-M_{\eta_1}^2) 
+ F_0^2 \delta(t) + \sum_{\pi'}^\infty F_{\pi'}^2 \delta(t-M_{\pi'})
\right] 
\ea
at {\em all} values of $t$. The
Weinberg-like Sum Rules, Eqs. (\ref{WSRS}),(\ref{weinSP1}),(\ref{weinSP2}),
assuming local duality holds above $s_0$, impose
\ba
{\dis \sum_V^{M_V^2<s_0}}\, f_V^2 \, M_V^2  -
{\dis \sum_A^{M_A^2<s_0}}\, f_A^2 \, M_A^2 
&=& F_0^2 \, ; \nonumber \\
{\dis \sum_V^{M_V^2<s_0}}\, f_V^2 \, M_V^4  -
{\dis \sum_A^{M_A^2<s_0}}\, f_A^2 \, M_A^4 
&=& 0 \, .
\ea
\ba
F_0^2+ {\dis \sum_{\pi'}^{M_{\pi'}^2<\tilde s_0}}\, F_{\pi'}^2&=&  
{\dis \sum_{\eta_1}^{M_{\eta_1}^2<\tilde s_0}}\, 
F_{\eta_1}^2  \, =
8 {\dis \sum_{S_8}^{M_{S_8}^2<\tilde s_0}}\, 
c_{m,S_8}^2  =
8 {\dis \sum_{S_1}^{M_{S_1}^2<\tilde s_0}}\, 
\tilde c_{m,S_1}^2  \, .
\ea

We will use the expression (\ref{LRNC})  to get one  of the two relevant
integrals defined in (\ref{ALRSP})  and the EM pion mass
difference in (\ref{EMpion})
\ba
{\cal A}_{LR}^{N_c}(\mu_R)&=&
2 {\dis \sum_V^{M_V^2<s_0}}\, f_V^2 \, M_V^6 \ln 
\left(\frac{M_V}{\mu_R} \right) -
2 {\dis \sum_A^{M_A^2<s_0}}\, f_A^2 \, M_A^6 \ln
\left(\frac{M_A}{\mu_R} \right)  \, ;
\nonumber \\
{\cal B}_{LR}^{N_c}&=&
2 {\dis \sum_V^{M_V^2<s_0}}\, f_V^2 \, M_V^4 
\ln  \left(\frac{M_V}{\mu_R} \right) -
2 {\dis \sum_A^{M_A^2<s_0}}\, f_A^2 \, M_A^4 
\ln \left(\frac{M_A}{\mu_R}  \right) 
\ea

We can also calculate the moments defined in (\ref{FESR})
\ba
M_{n+2}^{N_c}&=& {\dis \sum_V^{M_V^2<s_0}}\, f_V^2 \, M_V^{2(n+3)}  -
{\dis \sum_A^{M_A^2<s_0}}\, f_A^2 \, M_A^{2(n+3)} \, .  
\ea
These sum rules and higher moments were
studied in \cite{PPR01} with the  MHA
Ansatz\footnote{Which in this case means
that the spectral functions are saturated by
the pion pole, the first axial-vector and the first
rho vector resonances.}. The results obtained using
MHA are \cite{PPR01} 
\ba
M_2^{\rm MHA}&=&-(4.8 \pm 1.6) \cdot 10^{-3} \, {\rm GeV}^6 \, , \quad
M_3^{\rm MHA}=-(8.1 \pm 2.7) \cdot 10^{-3} \,  {\rm GeV}^8 \nonumber \\
{\cal A}_{LR}^{\rm MHA}(2 {\rm GeV})&=& 
(2.8\pm1.0) \cdot  10^{-3} \, {\rm GeV}^6 \, , \quad 
{\cal B}_{LR}^{MHA}=-(5.9\pm 2.0) \, \cdot 10^{-3} \, {\rm GeV}^4 \, .
\ea
for any value of $s_0$ larger than the 1st duality point,
i.e. $s_0 \geq 1.5$ GeV$^2$ by construction.

These should be compared with our results (\ref{data})
from the second duality point
or (\ref{data1st})
from the first duality point or the fit (\ref{ALEPH})\cite{DHGS98}.
The results using the MHA Ansatz are compatible with 
the ones using the first duality point 
within one sigma. 

If one uses the second duality point,
the moment $M_2$  agrees borderline within errors
but their central value is more than twice our result 
(\ref{data}). However the physical quantiy 
${\cal A}_{LR}(2 {\rm GeV})$
agrees within errors with our second duality point
though  our  central value is larger.  
The moment $M_3$ is more problematic and we
find that in the second duality point
it changes sign.

Also \cite{DHGS98} obtained the values of 
the condensates of dimension six an eight
which are proportional to $M_2$ and $M_3$ from a fit to
different type of moments of the same data.
The results are in Eq. (\ref{ALEPH}).
The moment $M_2$ is compatible within errors with our
results using the second duality point but not
with the our results using the first duality point. 
In particular, the  moment  $M_3$ is incompatible with  both 
our results using  the first and second duality points.
This must be due to the duality violations being weighted 
differently in (\ref{ALEPH}) as mentioned previously.

We conclude from these comparisons that
for the dimension eight moment $M_3$, local duality
at low values of $s_0$ starts to be a problem.
As we go up in the moments we need 
more and more accurate information at higher energies.
This affects also to $M_2$ .

The case of more than one resonance in each channel
 was analyzed in full generality in \cite{KR98}. 
There one can find the large differences
that  the $\rho'$ produces in the second, third moments,
and higher moments  with respect 
to the case of a single vector resonance in each channel.

\section{Conclusions}
\label{conclusions}

In this work, we have calculated in a model independent
way the matrix elements of the $\Delta S=1$ operators
$Q_7$ and $Q_8$ in the chiral limit. We have done it
to  all orders in $1/N_c$ and NLO in $\alpha_s$.

The scheme dependence has been taken into account
exactly at NLO using the $X$ boson method
as proposed and used in \cite{scheme,epsprime,deltaI}. In fact, these two
operators are a submatrix of the ten by ten done 
in \cite{epsprime}. 

We would like to mention some issues
sometimes mixed up in the literature.
First, the $X$ boson method has nothing to do with using
or not large $N_c$. It can be used without the large $N_c$
approximation as well, as shown again in this paper.
Second, our method of treating the scheme
dependence is consistent and we never 
mix up two different schemes, cut-off and $\overline{MS}$ schemes.
We do an analytic matching between a cut-off 
regularization and dimensional regularization
in a well-defined scheme at perturbative scales first.
The finite parts arising in this matching appear in the methods
using dimensional regularization to the end as well as explained in the
appendix.

We obtain exact matching in an Euclidean-cut-off regularization
and  analytical cancellation exact of (all) infrared and UV 
scheme dependences.

For the contribution of higher order operators
discussed in \cite{CDG00} and \cite{NAR01}
we clarify how to include all higher dimensional operators
and  exact scheme dependence at NLO in $\alpha_S$ of both 
the $Q_7$ and $Q_8$ matrix elements. As a result we find 
smaller corrections due to this effects as discussed 
in Section \ref{higherdim}. In our approach the effect 
of the higher order operators is to remove the
remaining dependence on the Euclidean cutoff $\mu$ beyond 
the RGE evolution. The result of resumming  all
higher dimensional operators in the case of
$Q_7$ makes its prediction much less sensitive
to the choice of $s_0$. 

As noticed in \cite{KPR01,epsprime}, ${\cal A}_{SP}$ is zero
in the large $N_c$ limit and therefore is Zweig suppressed.
We find  no sizeable violation of the dimension six FESR
using factorization for $Q_8$.

We find that the moment $M_2$ is very sensitive to the 
spectral function arount 2~GeV$^2$.

Our main analytical results are the expression for the
matrix-elements (\ref{imge}), the bag parameters (\ref{resbag1}),
(\ref{resbag2}) and the expansion coefficients of the 
spectral functions
(\ref{resPILR}), (\ref{F11}) and (\ref{resPISP}). The main 
numerical results
are the VEVs (\ref{numO1}),(\ref{numO2}) and the bag 
parameters (\ref{numbag}).
These results are exact in the chiral limit, so we have
the $\Delta I=3/2$ part of $\varepsilon'/ \varepsilon$
model independently at all orders in $1/N_c$.
In order to reach final values all effects which vanish 
in the chiral limit,
as final state interactions, quark-mass effects, isospin violation
and long-distance electromagnetic effects still need to be included.

\section*{Acknowledgements}
We thank Andreas H\"ocker for checking our calculations of $M_2$
and $M_3$ with the ALEPH data, Bachir Moussallam for the
programs used in \cite{MOU00}. 
We thank the authors of \cite{KPR01} and
\cite{CDGM01} for sending us their
draft previous to publication. We thank Vincenzo Cirigliano, 
John Donoghue, Santi Peris, Toni Pich and Eduardo de Rafael for
useful discussions.
This work is supported by the Swedish Research Council,
the European Union TMR network, Contract No. ERBFMRX--CT980169 
(EURODAPHNE), by MCYT (Spain), Grant No. FPA 2000-1558
and by Junta de Andaluc\'\i a, Grant No. FQM-101.
EG is indebted to MECD (Spain) for a FPU fellowship.
\appendix

\section{Calculation of the Corrections of $\mathcal{O}(a^2)$ to the
Dimension Six Contribution to $\Pi_{LR}^T(Q^2)$}
\label{AppA}
\subsection{Renormalization Group Analysis}

We have the two-point function
\begin{eqnarray}
\Pi^{\mu \nu}_{LR} (q)&\equiv& \frac{1}{2}\,
 i\int d^Dy\, e^{iq \cdot y}\langle 0\vert
T(L^{\mu}(y)R^{\nu}(0)^{\dag} )\vert 0\rangle \equiv
(q^{\mu}q^{\nu}-g^{\mu \nu} q^2)\Pi_{LR}^{T} (q^2) \nonumber \\ 
&+&q^{\mu} q^{\nu}\Pi_{LR}^{L}(q^2)
\end{eqnarray}
The contribution of dimension six operators to $\Pi_{LR}^{T}(Q^2)$ (where
$Q^2=-q^2$) can be written in $D=4-2\epsilon$ dimensions as
\begin{equation}\label{dimension6}
Q^6\, \Pi_{LR}^{T} (Q^2)\Big \vert_{D=6}
\,\equiv \, \nu^{2\epsilon} \, \sum_{i=1,2} C_i (\nu,Q^2) <O_i>(\nu)
\end{equation}
with
\begin{eqnarray}\label {operadores}
<O_1>&\equiv& \langle 0 | O_6^{(1)} | 0 \rangle =
\frac{1}{4} \,
 <0\vert(\overline s \gamma ^{\nu} d)_L (\overline
d\gamma_{\nu} s)_R\vert 0> \nonumber \\
<O_2>&\equiv& \langle 0 | O_6^{(2)} | 0 \rangle =
3 \, <0\vert(\overline d d)_L (\overline s s)_R\vert 0>
\end{eqnarray}
and
\begin{equation}
\label{Wilson}
C_i (\nu,s)=a(\nu)\sum_{k=0} a(\nu)^k\,  C_i ^{(k)} (\nu,s)
\end{equation}
where the dependence in $\nu$ and $s$ of $C_i^{(k)}$ is only
logarithmic. Everything here we define in the $\overline{MS}$ scheme.

In absence of electromagnetic interactions the matrix elements
(\ref{operadores}) only mix between themselves. The 
renormalization group equations (RGE) they satisfy are
\begin{eqnarray}\label{EGR}
\nu \frac{d<O_1> (\nu)}{d\nu} &=&-\gamma_{77}(\nu)<O_1>(\nu)+\frac{1}{6}\,
\gamma_{87}(\nu)<O_2>(\nu) \nonumber\\
\nu \frac{d<O_2> (\nu)}{d\nu} &=&-\gamma_{88}(\nu)<O_2>(\nu)+6\,
\gamma_{78}(\nu)<O_1>(\nu)
\end{eqnarray}
With $\gamma(\nu)$ the QCD anomalous dimension matrix
defined in (\ref{gamma}).
In the $NDR$ scheme \cite{BJLW93,CFMR94,BBL96}
\footnote{For these operators the Fierzed
version and the $Q_7$-$Q_8$ version have the same anomalous 
dimension matrix.} for $n_f=3$ flavours\footnote{We will use
along this work $n_f=3$ since this is the number of active flavours
of the QCD effective theory where $Q_7$ and $Q_8$ appear.}, 
\ba
\label{NDR}
\gamma(\nu)&=& {\dis  \sum_{n=1}} \gamma^{(n)} 
 a(\nu)^n  \, \nonumber \\
\gamma^{(1)}&=& -\frac{3}{2 \, N_c} \, \left( 
\begin{array}{c|c}
-1 & 0 \nonumber \\ N_c & N_c^2 -1 \nonumber \\ 
\end{array} \right) ;  \nonumber \\ 
\gamma^{NDR (2)} &=&   - \frac{1}{96 N_c^2} \left( 
\begin{array}{c|c}
-137 N_c^2 + 132 N_c -45       &  
213 N_c^3 - 72  N_c^2  +  108 N_c \nonumber \\ 
200 N_c^3  - 132  N_c^2 -18 N_c & 
203 N_c^4 - 60 N_c^3  - 479 N_c^2 + 132  N_c  - 45  
\end{array} \right)  \, . \nonumber \\ 
\ea
In the HV scheme of \cite{BJLW93,BBL96} \footnote{I.e. without
the $\beta_1 \, C_F  $ terms from renormalizing the axial current
in the diagonal coefficients \cite{BBL96}.}
\ba
\label{HV}
\gamma^{HV (2)} &=&   - \frac{1}{96 N_c^2} \left( 
\begin{array}{c|c}
 - 17 N_c^2 - 12 N_c - 45 &
-107 N_c^3 + 24 N_c^2 + 108 N_c \nonumber \\ 
80 N_c^3 + 12 N_c^2  - 18 N_c & 
115 N_c^4 - 12  N_c^3 - 71 N_c^2 - 12 N_c - 45 
\end{array} \right)  \, . \nonumber \\ 
\ea

We also need the quark mass anomalous dimension
in the $\overline{MS}$ scheme, 
\ba
\gamma_m(a)\equiv -\frac{\nu}{m} \, \frac{{\rm d} m}{{\rm d} \nu}
= {\dis \sum_{k=1}} \, \gamma_m^{(k)} a(\nu)^k \, 
\ea
where $m$ is a quark mass. 
The first coefficient is scheme independent
\be
\gamma_m^{(1)}= \frac{3}{2} C_F \, . 
\ee
Notice that $\gamma_{88}^{(1)}=- 2 \gamma_m^{(1)}$
to {\em all} orders in $1/N_c$ \cite{BG86,deR89},
this is the reason why $B_8$ in the chiral limit
is very near to 1 \cite{epsprime}.  The large $N_c$ result absorbs
{\em all} the one-loop scale dependence.
This exact scale cancellation {\em does} not occur
for $Q_6$ even at leading order
in $\alpha_S$. There is  a remnant diagonal anomalous dimension at one-loop
of order one  in $1/N_c$ which is not taken into account by
the large $N_c$ matrix element.
There is therefore no reason to expect $B_6$ around 1
as sometimes is claimed in the literature.

$\gamma_m^{(2)}$
is the same for both the NDR and HV schemes\cite{TAR81}, 
\ba
\gamma^{\overline {MS} (2)}_m&=&
\frac{C_F}{96 N_c} \left[203 N_c^2 - 60  N_c - 9\right] \, .
\ea
The relation $\gamma_{88}^{(2)}=- 2 \gamma_m^{(2)}$ is  not valid:
\ba
\gamma_{88}^{NDR (2)} &=& - 2 \gamma_m^{\overline{MS}(2)}
 +\frac{1}{32 N_c^2}
\left[ 89 N_c^2 - 24 N_c + 18 \right]  \,.
\ea 

The two-point function $\Pi_{LR}^T(Q^2)$ is  independent of
the scale $\nu$ in $D=4$
\begin{equation}\label{A7}
\frac{d}{d\nu}\Big(Q^6\, \Pi_{LR}^T (Q^2)\Big \vert_{D=6}\Big )=0 \, .
\end{equation}
This is also true in $D$ dimensions if 
$\gamma_5$ is anti-commuting like in the NDR scheme. 
The HV results are obtained from the NDR ones using 
the published results in  \cite{CFMR94}.

In $D=4-2\epsilon$ (\ref{A7})  yields the general condition
\ba
0&=&\dis \sum_{k=0} a^k(\nu) 
\left(  \beta(a) (k+1) - 2 \epsilon k) C_i^{(k)}(\nu,Q^2)\, <O_i>(\nu)
\right. \nonumber \\ 
&+& \left. \nu \frac{d C_i^{(k)}(\nu,Q^2)}{d \nu} <O_i>(\nu) 
+ C_i^{(k)}(\nu,Q^2) \, \nu \frac{d<O_i>(\nu)}{d\nu} \right)
\ea
with
\be
\nu \frac{da(\nu)}{d\nu} = a \, (\beta(a) - 2\epsilon)
\ee
and
$\beta(a)= \sum_{k=1} \, \beta_{k} a(\nu)^{k} $
with  first coefficient $\beta_1=1- 11 N_c/6$ for $n_f=3$.

To order $a(\nu)^0$,  one gets 
\begin{equation}
\frac{dC_i^{(0)}(\nu,Q^2)}{d\nu}=0
\end{equation}
so the $C_i^{(0)}$ are constants.

To order $a(\nu)$
\begin{eqnarray}
&& \beta_1 C_1^{(0)}+\nu
\frac{dC_1^{(1)}(\nu,Q^2)}{d\nu}- \gamma_{77}^{(1)} C_1^{(0)}
-2 \epsilon C_1^{(1)}=0 \, , \nonumber \\
&&\beta_1 C_2^{(0)}+  \nu
\frac{dC_2^{(1)}(\nu,Q^2)}{d\nu}- \gamma_{88}^{(1)}C_2^{(0)}
+\frac{1}{6} \gamma_{87}^{(1)} C_1^{(0)} 
-2 \epsilon C_2^{(1)}=0\, .
\end{eqnarray}
Integrating these two equations we obtain
\begin{eqnarray}
\label{C11}
C_1^{(1)} (\nu,Q^2)&=&
\frac{D_1^{(1)}}{2 \epsilon}+
\left(\frac{Q^2}{\nu^2}\right)^{- \epsilon} 
\left[ - \frac{{D_1^{(1)}}}{2\epsilon} + F_1^{(1)}\right]\, ,
\nonumber \\ 
C_2^{(1)} (\nu,Q^2)&=&
\frac{D_2^{(1)}}{2 \epsilon}+
\left(\frac{Q^2}{\nu^2}\right)^{- \epsilon} 
\left[ - \frac{{D_2^{(1)}}}{2\epsilon} + F_2^{(1)}\right]
\end{eqnarray}
with
\begin{equation}
\label{D11}
D_1^{(1)}=C_1^{(0)} \left[
\beta_1 - \gamma_{77}^{(1)} \right] \, ; \hspace{1 cm}
D_2^{(1)}=C_2^{(0)} \left[ 
\beta_1 -\gamma_{88}^{(1)} \right] +
\frac{1}{6} C_1^{(0)} \, \gamma_{87}^{(1)}
\end{equation}
which are valid in $D=4-2\epsilon$. 
The coefficients $C_i^{(0)}$,
$D_i^{(1)}$, and $F_i^{(1)}$ depend on $\epsilon$.
The anomalous dimensions $\beta_1$
and $\gamma_{ij}$ do not depend on $\epsilon$ in $MS$,
and in $\overline{MS}$ schemes in a known fashion.

\subsection{Calculation of the Constants $C_i^{(0)}$ and $F_i^{(1)}$}
\label{appCi}

The bare vacuum expectation value of $<O_1>$ can be expressed as an
integral as follows
\begin{equation} 
\label{O1integral}
<O_1>^{\rm bare}=-\frac{i}{2}g_{\mu \nu}\int
\frac{d^Dq}{(2\pi)^D}\Pi^{\mu\nu}_{LR}(q)=
\frac{D-1}{2}\int\frac{d^DQ}{(2\pi)^D} (Q^2\Pi_{LR}^T(Q^2)) \, . 
\end{equation}
The scheme used here to regularize this integral is 
the  $\overline{MS}$ scheme with $D=4-2\epsilon$, 
\begin{equation}\label{A2}
<O_1>^{\rm bare}=\frac{3-2\epsilon}
{32\pi^2}\frac{(4\pi)^{\epsilon}}
{\Gamma(2-\epsilon)}\int _0^{\infty} dQ^2 (Q^2)^{1-\epsilon}
(Q^2 \Pi_{LR}^T(Q^2))
\end{equation}
Notice that $<O_1>^{\rm bare}$  is scale independent.
The integral (\ref{A2}) diverges due to the high energy behaviour of
$\Pi_{LR}^T(Q^2)$. It is enough then to use the large $Q^2$
expansion of $\Pi_{LR}^T(Q^2)$ in $D$ dimensions.
This is a series in $\left(1/Q^2\right)^n$ starting at
$n=3$ in the chiral limit, Eq. (\ref{dimension6}).
Each coefficient of this series is finite
and can be written as a Wilson coefficient times the vacuum 
expectation value of some operator.
We now put (\ref{dimension6}) and (\ref{C11})
in (\ref{A2}) and perform the integral to find the divergent part.
For that we need the integral,
\begin{eqnarray}
\int_{\mu^2}^{\infty}
dQ^2\frac{1}{(Q^2)^{1+\epsilon}}&=&\frac{1}
{\epsilon}\mu^{-2\epsilon}\,.
\end{eqnarray}
We will set $\mu=\nu$ afterwards.

The $\overline{MS}$ subtraction needed then gives the full 
dependence on $\nu$.
\begin{eqnarray}
\label{o1run}
\nu\frac{d<O_1>^{\overline{MS}}(\nu)}{d\nu}
&=&\frac{3}{16\pi^2}a(\nu)\left\lbrace
\overline C_1^{(0)} + a(\nu)\left\lbrace  
\frac{G_1^{(1)}}{2} + \frac{\overline D_1^{(1)}}{6} + \overline F_1^{(1)} 
\right\rbrace\right\rbrace<O_1>^{\overline{MS}}(\nu)\nonumber \\
&+&\frac{3}{16\pi^2}a(\nu) \left\lbrace \overline C_2^{(0)} +a(\nu) 
\left\lbrace 
\frac{G_2^{(1)}}{2} + \frac{\overline D_2^{(1)}}{6}+\overline F_2^{(1)}
\right\rbrace\right\rbrace<O_2>^{\overline{MS}}(\nu) \, .\nonumber \\
\end{eqnarray}
Overlined quantities are in four dimensions and
\ba
G_i^{(1)} &=& \lim_{\epsilon\to 0}\frac{1}{\epsilon}
 \left(D_i^{(1)}-\overline D_i^{(1)}\right)\,.
\ea
Comparing (\ref{o1run}) and  (\ref{EGR}) 
order by order in $a$ and using (\ref{D11}), we get
up to the needed order in $\epsilon$
\ba
\label{resPILR}
C_1^{(0)}&=& -\frac{16\pi^2}{3} 
\left[ \gamma_{77}^{(1)}+ p_{77} \, \epsilon \right]
\,;
%\quad\quad
%D_1^{(1)}=
%-\frac{16\pi^2}{3}
%\left[ \gamma_{77}^{(1)}+ p_{77} \, \epsilon \right]
%\left[ \beta_1  - \gamma_{77}^{(1)} \right] \,; 
 \nonumber \\
C_2^{(0)}&=&\frac{8 \pi^2}{9}
\left[ \gamma_{87}^{(1)} + p_{87} \, \epsilon \right];
% \,;\quad
%D_2^{(1)}= 
%\frac{8\pi^2}{9} 
%\left[ \left(\gamma_{87}^{(1)}+ p_{87} \, \epsilon \right)
%\left( \beta_1  -\gamma_{88}^{(1)} \right)- 
%p_{77} \gamma_{87}^{(1)}\, \epsilon  \right] \, ; 
%\nonumber \\
%G_2^{(1)}&=&\frac{8\pi^2}{9}\, \left[
%p_{87} \,  \left[ \beta_1 - \gamma_{88}^{(1)}\right] 
%- p_{77}\,  \gamma_{87}^{(1)}\right]\, ; \quad \quad
%G_1^{(1)}=-\frac{16\pi^2}{3}\, 
%p_{77}\,  \left[ \beta_1 - \gamma_{77}^{(1)}\right] 
\ea
$\overline{D_1^{(1)}}$, $\overline{D_2^{(1)}}$, $G_1^{(1)}$
and   $G_2^{(1)}$ are then determined up to the $p_{ij}$ from Eq. (\ref{D11}).
We also get
\ba
\label{F11}
\overline{F}_1^{(1)}&=& - \frac{16 \pi^2}{3} \gamma_{77}^{(2)}-
\frac{1}{6}\overline{D}_1^{(1)}-\frac{1}{2}G_1^{(1)}\,;
\nonumber\\
\overline{F}_2^{(1)}&=& \frac{ 8 \pi^2}{9 } \gamma_{87}^{(2)}
-\frac{1}{6}\overline{D}_2^{(1)}-\frac{1}{2}G_2^{(1)}\,.
\ea
The constants $p_{ij}$ we determine below.

\subsection{The constants $p_{ij}$}

We now evaluate Eq. (\ref{A2}) to ${\cal O}(a)$ fully with its subtraction
in dimensional regularization using the same split in the integral
at $\mu^2$ as we used in the main text. The short-distance dimension
six part is the only divergent part, now regulated by dimensional
regularization rather than the $X$-boson propagator as in the main text.
The result is
\ba
<O_1>^{\overline{MS}}(\nu) &=&
\frac{3}{32\pi^2}a(\nu)\Bigg[\left(\frac{1}{3}\overline{C}_1^{(0)}
-\frac{16\pi^2}{3}p_{77}\right)<O_1>^{\overline{MS}}(\nu)
\nonumber\\
&+&\left(\frac{1}{3}\overline{C}_2^{(0)}
+\frac{8\pi^2}{9}p_{87}\right)<O_2>^{\overline{MS}}(\nu)\Bigg]
-\frac{3}{32\pi^2}{\cal A}_{LR}(\nu).
\ea
Comparison with Eq. (\ref{numresult}) and (\ref{numresultHV})
allows to determine $p_{77}$ and $p_{87}$. The finite coefficients
there are basically the $\Delta r_{ij}$ that corrected for the
dimensional regularization to the $X$-boson scheme. If one works
fully in dimensional regularization, it is here that these finite parts
surface.

The result is
\ba
p_{77}^{NDR}&=&-\frac{3}{4 N_c}\;; \quad 
p_{87}^{NDR}=\frac{3}{4}\;;
\nonumber\\
p_{77}^{HV}&=&-\frac{9}{4 N_c}\;; \quad 
p_{87}^{HV}=\frac{9}{4} \, .
\ea 
The transition between both agrees with the results in \cite{CFMR94}.

Putting in (\ref{NDR}) and (\ref{HV}) to obtain numerical values
\ba
-\frac{3}{16\pi^2}\overline{C}_1^{(0)}&=&
%\frac{3}{2 N_c} =
 \frac{1}{2}\, ;
\quad\quad
-\frac{3}{16\pi^2}\overline{D}_1^{(1)}
 =  -\frac{5}{2} \, ;
\nonumber \\
\frac{9}{8\pi^2}\overline{C}_2^{(0)}&=-&\frac{3}{2}  ;
\quad\quad
\frac{9}{8\pi^2} \, 
\overline{D}_2^{(1)}
%&=& \frac{1}{2 N_c}[N_c^2- 3 N_c+9]
=\frac{3}{2} \, .
\ea
For (\ref{F11}), in the NDR case we get
\ba
-\frac{3}{16\pi^2}\overline{F}_1^{NDR(1)} &=&
% \frac{1}{96 N_c^2} \left[ 181 N_c^2 
% - 156  N_c + 81 \right] = 
\frac{13}{16}\, \quad\quad
 \frac{9}{8\pi^2}\overline{F}_2^{NDR(1)} 
% &=& \frac{-1}{48 N_c} \left[ 104 N_c^2 
% - 78 N_c + 27 \right]
  = -\frac{75}{16} 
\ea
and in the HV scheme
\ba
-\frac{3}{16\pi^2} \overline{F}_1^{HV(1)} &=& 
% -\frac{3}{16\pi^2} F_1^{NDR(1)} + \frac{3}{2 N_c } \beta_1 
% + \frac{3}{2}  =
 \frac{41}{16}\,; \quad\quad
\frac{9}{8\pi^2}\overline{F}_2^{HV(1)} 
% &=&  \frac{9}{8\pi^2} F_2^{NDR(1)} - \frac{3}{2} \beta_1 
%- \frac{3}{2} N_c   
= - \frac{63}{16}\,.
\ea
All the expressions above are for $n_f=3$ flavours.

\section{Calculation of the Corrections of $\mathcal{O}(a^2)$ to the
Dimension Six Contribution to $\Pi_{SS+PP}^{(0-3)\, \rm conn}(Q^2)$}
\label{AppB}

\subsection{Renormalization Group Analysis}

The function we have to study here is 
\be
\label{B1p}
\Pi_{SS+PP}^{(0-3)}(q)\equiv i\int\,d^Dy \, e^{iy \cdot q}\,
\langle 0\vert T[(S+iP)^{(0-3)}(y)(S-iP)^{(0-3)}(0)]\vert 0 \rangle
\ee
with the definitions appearing in Section \ref{Q7Q8}.

The contribution of dimension six to the connected part of 
$\Pi_{SS+PP}^{(0-3)}(Q^2)$ can be written as (\ref{SVZSP})
\be\label{expsspp}
Q^4\,\Pi_{SS+PP}^{(0-3)\, \rm conn}(Q^2)\Big \vert _{D=6}
= \nu^{2 \epsilon}
\sum_{i=1,2} \tilde C_i(\nu,Q^2)  \langle O_i \rangle(\nu) \, 
\ee
with
\begin{equation}
\label{SSWilson}
\tilde C_i (\nu,s)=a(\nu)\sum_{k=0} a^{k}(\nu) \tilde C_i ^{(k)} (\nu,s)
\end{equation}
and the operators $O_1$  and $O_2$ were 
defined in (\ref{operadores}). 

From (\ref{scaleSS}) and (\ref{dispersionSS}),
we have now  in $D=4-2\epsilon$, 
\be
\nu \frac{d}{d\nu}\Pi_{SS+PP}^{(0-3)}(Q^2)=2\gamma_m(\nu)
\, \Pi_{SS+PP}^{(0-3)}(Q^2) \, .
\ee
Using this relation and the renormalization group equations, 
we get
\begin{eqnarray}
\tilde C_1^{(1)} (\nu,Q^2)&=&
\frac{\tilde D_1^{(1)}}{2 \epsilon}+
\left(\frac{Q^2}{\nu^2}\right)^{- \epsilon} 
\left[ - \frac{{\tilde D_1^{(1)}}}{2\epsilon} 
+ \tilde F_1^{(1)}\right]\, ,
\nonumber \\ 
\tilde C_2^{(1)} (\nu,Q^2)&=&
\frac{\tilde D_2^{(1)}}{2 \epsilon}+
\left(\frac{Q^2}{\nu^2}\right)^{- \epsilon} 
\left[ - \frac{{\tilde D_2^{(1)}}}{2\epsilon} 
+ \tilde F_2^{(1)}\right]
\end{eqnarray}
with
\ba
\label{D11tilde}
\tilde D_1^{(1)}&=&\tilde C_1^{(0)}\left[ 
\beta_1- 2\gamma_m^{(1)} - \gamma_{77}^{(1)} \right] 
 \, ,\nonumber\\
\tilde D_2^{(1)}&=&\tilde C_2^{(0)} 
\left[ \beta_1 - 2\gamma_m^{(1)}- \gamma_{88}^{(1)} \right]
+ \frac{1}{6} \tilde C_1^{(0)} \, \gamma_{87}^{(1)} \, .
\ea

In the next Section we determine the values of the constants
 $\tilde C_i^{(0)}$ and  $\tilde F_i^{(1)}$, 
which depend on $\epsilon$.

\subsection{Calculation of the Constants 
$\tilde C_i^{(0)}$ and $\tilde F_i^{(1)}$}
\label{AppA2}

The connected part of 
$\Pi_{SS+PP}^{(0-3)}(Q^2)$ can be related to the bare vacuum 
expectation value of the connected part of 
$<O_2>(\nu)$ through the relation \be\label{o2}
<O_2>^{\rm bare}_{\rm conn}(\nu)
=-i\,\int\frac{d^Dq}{(2\pi)^D}\,\Pi_{SS+PP}^{(0-3) \, \rm conn}(q)=
\int\frac{d^DQ}{(2\pi)^D}\,\Pi_{SS+PP}^{(0-3) \, \rm conn}(Q^2)
\ee
In the $\overline {MS}$ scheme with $D=4-2\epsilon$ and with renormalized 
$\Pi_{SS+PP}^{(0-3)}(Q^2)$
\be\label{O2MS}
<O_2>^{\rm bare}_{\rm conn}(\nu)
=\frac{(4\pi)^{\epsilon}}
{16 \pi^2 \, \Gamma(2-\epsilon)}\,\int _0 ^{\infty} \, dQ^2\,
(Q^2)^{1-\epsilon}\,  \Pi_{SS+PP}^{(0-3)\, \rm conn}(Q^2)
\ee

Proceeding analogously to the case of $\Pi_{LR}^T(Q^2)$
in Appendix \ref{appCi} and using that 
there is now a non vanishing  contribution coming 
from the anomalous dimensions of $\Pi_{SS+PP}^{(0-3)\, \rm conn}$, 
namely, 
\be
\int\frac{d^DQ}{(2\pi)^D}\,
\nu \, \frac{d \Pi_{SS+PP}^{(0-3)\, \rm conn}(Q^2)}{d\nu}
=2\gamma_m(\nu) <O_2>^{\rm bare}_{\rm conn}(\nu)
\ee
that we have to add to the one from the $\nu$-dependence
of the subtraction determined by the integration of $Q^2$ in 
(\ref{o2}).

The scale dependence  of  
the total $<O_2>$ can be obtained by adding both, we get 
in $D=4-2\epsilon$
\begin{eqnarray}
\nu\frac{d<O_2>^{\overline{MS}}(\nu)}{d\nu}
&=&\frac{1}{8\pi^2}a(\nu)\left\lbrace
\overline{\tilde C}_1^{(0)}+a(\nu)\left\lbrace 
\frac{\tilde G_1^{(1)}}{2} + \frac{\overline{\tilde D}_1^{(1)}}{2} +
\overline{\tilde F}_1^{(1)}
\right\rbrace \right\rbrace<O_1>^{\overline{MS}}(\nu)\nonumber \\
&+&\frac{1}{8\pi^2}a(\nu)\left\lbrace \overline{\tilde C}_2^{(0)}+
a(\nu)\left\lbrace  \frac{\tilde G_2^{(1)}}{2}
+ \frac{ \overline{\tilde D}_2^{(1)}}{2}
+ \overline{\tilde F}_2^{(1)}\right\rbrace \right \rbrace
<O_2>^{\overline{MS}}(\nu) \nonumber \\
&+&2\,\gamma_m <O_2>^{\overline{MS}}(\nu)
\end{eqnarray}
Again the barred quantities have to be taken at  $\epsilon=0$ and
\ba
\tilde G_i^{(1)} = \lim_{\epsilon\to0}
\frac{1}{\epsilon}\left(\tilde D_i^{(1)}-\overline{\tilde D}_i^{(1)}\right)\,.
\ea

Comparing this equation with (\ref{EGR}) order by order in $a(\nu)$ one 
obtains,
\begin{eqnarray}
\label{resPISP}
\tilde C_1^{(0)}&=& 48 \pi^2  p_{78}\, \epsilon \, ;
 \quad\quad\quad\quad
\tilde C_2^{(0)}= -8 \pi^2 \, 
\left[ \gamma_{88}^{(1)}+ 2 \gamma_m^{(1)}+  
p_{88} \, \epsilon \right] \, ;
\nonumber \\ 
%\tilde D_1^{(1)}&=& 48 \pi^2 p_{78}
%\left[ \beta_1 -  2 \gamma_m^{(1)} -\gamma_{77}^{(1)}\right] 
%\, \epsilon \, ; \nonumber \\ 
% \tilde D_2^{(1)} &=& - 8 \pi^2 
%\left[ \left(\beta_1 -  2 \gamma_m^{(1)} -\gamma_{77}^{(1)}\right) 
%\left(\gamma_{88}^{(1)}+2\gamma_m^{(1)}+ p_{88}\epsilon \right) 
%- p_{78} \gamma_{87}^{(1)} \, \epsilon \right] \, ; 
%\nonumber \\
%\tilde G_1^{(1)}&=& 48 \pi^2 p_{78} 
%\left(\beta_1 - 2\gamma_m^{(1)} - \gamma_{77}^{(1)}\right) \, ;  
%\nonumber \\
%\tilde G_2^{(1)}&=& -8 \pi^2 \left[ p_{88} 
%\left(\beta_1 - 2\gamma_m^{(1)} - \gamma_{88}^{(1)}\right) 
%- p_{78} \gamma_{87}^{(1)} \right] \, ;  
%\nonumber \\
\overline{\tilde F}_1^{(1)}&=& 48\pi^2 \gamma_{78}^{(2)}
-\frac{1}{2}\tilde G_1^{(1)}
-\frac{1}{2}\overline{\tilde D}_1^{(1)}
\, ;\nonumber \\
\overline{\tilde F}_2^{(1)}&=& -8 \pi^2
\left[ \gamma_{88}^{(2)}+2\gamma_m^{(2)}\right]
-\frac{1}{2}\tilde G_2^{(1)}
-\frac{1}{2}\overline{\tilde D}_2^{(1)}
\end{eqnarray}
Using Eq. (\ref{D11tilde}) everything can then be determined in terms of
the $p_{ij}$.

\subsection{Calculation of the $p_{ij}$.}

We now evaluate also the finite part from Eq. (\ref{B1p}) fully in dimensional
regularization to ${\cal O}(a)$ and obtain
\ba
<O_2>^{\overline{MS}}(\nu) &=&
\frac{1}{16\pi^2}a(\nu)\Bigg[\left(\overline{\tilde C}_1^{(0)}
+{48\pi^2}p_{78}\right)<O_1>^{\overline{MS}}(\nu)
\nonumber\\
&+&\left(\overline{\tilde C}_2^{(0)}
-{8\pi^2} p_{88}\right)<O_2>^{\overline{MS}}(\nu)\Bigg]
+3<0|\bar qq|0>^2(\nu)
\nonumber\\
&+&\frac{1}{16\pi^2}{\cal A}_{SP}(\nu)\,.
\ea
Comparison with Eq. (\ref{O6operator})
allows to determine $p_{78}$ and $p_{88}$. The finite coefficients
there are basically the $\Delta r_{ij}$ that corrected for the
dimensional regularization to the $X$-boson scheme. If one works
fully in dimensional regularization, it is here that these finite parts
surface for the $Q_8$ contribution.

The results are
\ba
p_{88}^{NDR}&=&-\frac{5}{4} N_c - \frac{1}{4 N_c}\;; \quad 
p_{78}^{NDR}=\frac{3}{2} \;;
\nonumber\\
p_{88}^{HV}&=&-\frac{9}{4} N_c +\frac{11}{4 N_c}\;; \quad 
p_{78}^{HV}=-\frac{1}{2} \, .
\ea 
again agreeing with the transition between both from 
\cite{CFMR94}.

Putting numbers, we get
\ba
\overline{\tilde C}_1^{(0)} &=& 
\overline{\tilde C}_2^{(0)} =
\overline{\tilde D}_1^{(1)} =
\overline{\tilde D}_2^{(1)} = 0\;
\nonumber\\
\frac{1}{48\pi^2} \overline{\tilde F}_1^{(1)} 
&=&  \frac{15}{32}\,; \quad\quad
 -\frac{1}{8\pi^2}\overline{\tilde F}_2^{(1)}
  = -\frac{211}{32}\;
\ea
which are scheme independent.
All the expressions above are for $n_f=3$ flavours.

\end{document}